\documentclass[twocolumn,amsmath,amssymb,floatfix,pre,floatfix]{revtex4} 

\usepackage{graphicx,color}
\definecolor{brown}{rgb}{0.63,0.27,0.18}
\definecolor{orange}{rgb}{1.00,0.65,0.00}

\usepackage{moresize}
\usepackage{dcolumn}
\usepackage{bm,multirow}

\begin{document}

\newcommand {\rsq}[1]{\langle R^2 (#1)\rangle}
\newcommand {\rsqL}{\langle R^2 (L) \rangle}
\newcommand {\rsqbp}{\langle R^2 (N_{bp}) \rangle}
\newcommand {\Nbp}{N_{bp}}
\newcommand {\etal}{{\em et al.}}
\newcommand{\Ham}{{\cal H}}
\newcommand{\RalfNew}[1]{\textcolor{red}{#1}}
\newcommand{\scs}{\ssmall}



\title{Beyond Flory theory: Distribution functions for interacting lattice trees}

\author{Angelo Rosa}
\email{anrosa@sissa.it}
\affiliation{
SISSA - Scuola Internazionale Superiore di Studi Avanzati, Via Bonomea 265, 34136 Trieste, Italy
}
\author{Ralf Everaers}
\email{ralf.everaers@ens-lyon.fr}
\affiliation{
Univ Lyon, Ens de Lyon, Univ Claude Bernard Lyon 1, CNRS, Laboratoire de Physique and Centre Blaise Pascal, F-69342 Lyon, France
}

\date{\today}

\begin{abstract}
While Flory theories~\cite{IsaacsonLubensky,DaoudJoanny1981,GutinGrosberg93,GrosbergSoftMatter2014,Everaers2016a} provide an extremely useful framework for understanding the behavior of interacting, randomly branching polymers, the approach is inherently limited.
Here we use a combination of scaling arguments and computer simulations to go beyond a Gaussian description.
We analyse distributions functions for a wide variety of quantities characterising the tree connectivities and conformations  
for the four different statistical ensembles, which we have studied numerically in Refs.~\cite{Rosa2016a,Rosa2016b}:
(a) ideal randomly branching polymers,
(b) $2d$ and $3d$ melts of interacting randomly branching polymers,
(c) $3d$ self-avoiding trees with annealed connectivity and
(d) $3d$ self-avoiding trees with quenched ideal connectivity.
In particular, we investigate the distributions 
(i) $p_N(n)$ of the weight, $n$, of branches cut from trees of mass $N$ by severing randomly chosen bonds;
(ii) $p_N(l)$ of the contour distances, $l$, between monomers;
(iii) $p_N(\vec r)$ of spatial distances, $\vec r$, between monomers, and 
(iv) $p_N(\vec r|l)$ of the end-to-end distance of paths of length $l$. 
Data for different tree sizes superimpose,  when expressed as functions of suitably rescaled observables $\vec x = \vec r/\langle r^2(N) \rangle$ or $x =l/\langle l(N) \rangle$. 
In particular, we observe a generalised Kramers relation for the branch weight distributions (i) and
find that all the other distributions (ii-iv) are of Redner-des Cloizeaux type, $q(\vec x) =  C \, |x|^\theta\  \exp \left( -(K |x|)^t \right)$. 
We propose a coherent framework, including generalised Fisher-Pincus relations, relating most of the RdC exponents to each other and to the contact and Flory exponents for interacting trees.
\end{abstract}

\maketitle

\section{Introduction}\label{sec:intro}
A randomly branched tree is a finite connected set of bonds that contain no closed loops and which is embedded in a $d$ dimensional space.
Aside from their importance in statistical physics~\cite{vanRensburgBook2015} and the intriguing connection to relevant physical problems as percolation~\cite{HavlinPRL1995},
randomly branched trees have received particular attention for being practically implicated in the modelling of branched~\cite{RubinsteinColby}, ring~\cite{KhokhlovNechaev85,RubinsteinRepton1987,RubinsteinPRL1994,RosaEveraersPRL2014,GrosbergSoftMatter2014,SmrekGrosberg2015,RubinsteinMacromolecules2016}
and supercoiled~\cite{MarkoSiggiaSuperCoiledDNA} polymers.

As customary in polymer physics~\cite{DoiEdwards,RubinsteinColby}, the behavior of randomly branched trees can be analyzed in terms of a small set of exponents describing how expectation values for observables characterising tree connectivities and conformations vary with the weight, $N$, of the trees or the contour distance, $L$, between nodes:
\begin{eqnarray}
\langle N_{br}(N) \rangle &\sim& N^{\epsilon}    \label{eq:epsilon}\\
\langle L(N) \rangle &\sim& N^\rho                     \label{eq:rho}\\
\langle R^2(L) \rangle &\sim& L^{2\nu_{path}}   \label{eq:nu_path}\\
\langle R_g^2(N) \rangle &\sim& N^{2\nu}         \label{eq:nu}\\
\langle N_c(N) \rangle &\sim& N^{\gamma_c}   \label{eq:gamma_c}\\
\langle N_c^{inter}(N) \rangle &\sim& N^{\beta} \label{eq:beta}
\end{eqnarray}
Here, $\langle N_{br}(N)\rangle$ denotes the average branch weight; 
$\langle L(N) \rangle$ the average contour distance or length of paths on the tree; 
$\langle R^2(L) \rangle$ the mean-square spatial distance between nodes with fixed contour distance; 
$\langle R_g^2(N) \rangle$ the mean-square gyration radius of the trees; 
$\langle N_c(N) \rangle$ the average number of intra-tree pair contacts;
$\langle N_c^{inter}(N) \rangle$ the average number of pair contacts between different trees in melt.
By construction, $\nu=\rho\, \nu_{path}$, and the relation $\epsilon=\rho$ is expected to hold in general~\cite{MadrasJPhysA1992}.

Exact values for the exponents are known only for a very few number of cases.
For ideal non-interacting trees, the exponents $\rho^{ideal}=\epsilon^{ideal}=\nu_{path}^{ideal}=1/2$ and $\nu^{ideal}=1/4$~\cite{ZimmStockmayer49,DeGennes1968}.
Furthermore, $\gamma_c^{ideal}=2-d \nu^{ideal}$ where $d$ denotes the dimension of the embedding space.
For interacting trees, the only known exact result~\cite{ParisiSourlasPRL1981} is the value $\nu=1/2$ for self-avoiding trees in $d=3$.
On the other hand,
numerical results~\cite{MadrasJPhysA1992,GrassbergerJPhysA2005,Rosa2016a,Rosa2016b} as well as approximate theoretical calculations~\cite{DerridaDeseze1982,JanssenStenullPRE2011} confirm
that Flory theories~\cite{IsaacsonLubensky,DaoudJoanny1981,GutinGrosberg93,GrosbergSoftMatter2014,Everaers2016a}
provide a useful framework for discussing the {\it average} behavior, Eqs.~(\ref{eq:epsilon}) to (\ref{eq:nu}), of a wide range of interacting tree systems.
While being remarkably successful though, due to its simplicity Flory theory is inevitably affected by serious known shortcomings and limitations~\cite{DeGennesBook,DesCloizeauxBook}.
They are, for instance, manifest in (small) deviations of predicted from observed or exactly know values for exponents as those defined in Eqs.~(\ref{eq:epsilon}) to (\ref{eq:gamma_c}),
and, importantly, it does not give any insight for the equally relevant exponent $\beta$, Eq.~(\ref{eq:beta}).
In the case of linear chains, these limitations are much more pronounced in the distribution functions for the corresponding observables which contain a wealth of additional information.

Little seems to be known about the even larger range of configurational distribution functions for interacting trees. 
In the following, we present numerical results based on a detailed analysis of five different tree ensembles, which we have simulated in \cite{Rosa2016a,Rosa2016b}:
(i) ideal randomly branching polymers,
(ii) $2d$ and $3d$ melts of interacting randomly branching polymers,
(iii) $3d$ self-avoiding trees with annealed connectivity and
(iv) $3d$ self-avoiding trees with quenched ideal connectivity.
We analyse the distributions along the lines of known relations for ideal trees and linear chains and propose a coherent framework, including generalised Fisher-Pincus relations, relating most of the exponents characterising the distribution functions to each other and to the contact and Flory exponents for interacting trees.

The paper is organised as follows:
In Section~\ref{sec:theory} we briefly summarise the theoretical background and list a number of useful results for ideal and interacting linear chains and trees.
In Section~\ref{sec:modmethods} we give a few details on the numerical methodologies employed for simulating the trees and analysing their connectivity.
Finally, we present and discuss our results in Sec.~\ref{sec:results} and briefly conclude in Sec.~\ref{sec:concls}.

\section{Theoretical background}\label{sec:theory}

\subsection{Ideal trees}
Consider a tree of mass $N$ defined as $N+1$ monomers connected by $N$ bonds or Kuhn segments. For ideal trees, Daoud and Joanny~\cite{DaoudJoanny1981} calculated the partition function ${\mathcal Z}_N$ in the continuum approximation:
\begin{equation}\label{eq:DJpartfunct}
{\mathcal Z}_N = \frac{I_1 \left( 2 \, \lambda \, N \right) }{\lambda \, N} \simeq
\left\{
\begin{array}{cc}
\frac{e^{2 \lambda N}}{2 \pi^{1/2} (\lambda N)^{3/2}}, \, & \lambda N \gg 1 \\
\\
1 + \frac{(\lambda N)^2}{2} , \, & \lambda N \ll 1
\end{array}
\right.
\end{equation}
where $I_1(x)$ is the first modified Bessel function of the first kind, $\lambda$ the branching fugacity and ${\mathcal Z}_0=1$. Removing a randomly chosen bond, splits a branch of size $n<N/2$ from the remaining tree of size $N-n-1$. For ideal trees, the probability distribution, $p_N(n)$, of branch sizes is given by the Kramers theorem~\cite{RubinsteinColby}
\begin{equation}\label{eq:Kramers_SplitProb}
p_N(n) = \frac{ {\mathcal Z}_n \, {\mathcal Z}_{N-1-n} } { \sum_{n=0}^{N-1} \, {\mathcal Z}_n \, {\mathcal Z}_{N-1-n} } \, ,
\end{equation}
and related to sum 
over all possible ways of splitting the tree.
From this expression, it is possible to derive the following asymptotic relations: 
\begin{equation}\label{eq:DaoudJoanny_SplitProb}
p_N(n) \simeq \left\{
\begin{array}{cc}
\frac{\lambda \, \left( \lambda N \right)^{3/2}}{4 \, \pi^{1/2} \left( \lambda n \right)^{3/2} \left( \lambda (N-n) \right)^{3/2}} , \, & \lambda n \gg 1\\
\\
\frac{\lambda}{4 \, \pi^{1/2} \left( \lambda n \right)^{3/2}} , \, & \lambda N \gg \lambda n \gg 1\\
\\
\frac{1}{N} , \, & \lambda N \ll 1
\end{array}
\right. \, ,
\end{equation}
with $\langle N_{br}(N) \rangle \sim N^{1/2} \equiv N^{2-x}$, where $x=3/2$ is the scaling exponent describing the decay of $p_N(n)$ in the large-$N$ limit, Eq.~(\ref{eq:DaoudJoanny_SplitProb}).

\subsection{Flory theory of interacting trees}
Flory theories are formulated as a balance of an entropic elastic term and an interaction energy~\cite{FloryChemBook}:
\begin{equation}\label{eq:fFlory}
{\mathcal F} = {\mathcal F_{el}(N,R)}+{\mathcal F_{inter}(N,R)} \, .
\end{equation}
In the standard case of self-avoiding walks,
$\frac{\mathcal F_{el}(N,R)}{k_BT} \sim \frac{R^2}{l_K^2 N}$
represents the entropic elasticity of a linear chain,
while
$\frac{\mathcal F_{inter}(N,R)}{k_BT} \sim v_2 \frac{N^2}{R^d}$
represents the two-body repulsion between segments, which dominates in good solvent. 
For interacting trees, the elastic free energy takes the form~\cite{GutinGrosberg93}: 
\begin{equation}\label{eq:fGutinPrefactors}
\frac{{\mathcal F}_{el}}{k_B T} \sim \frac{R^2}{l_K L} +  \frac{L^2}{N l_K^2} \, . 
\end{equation}
The expression reduces to the entropic elasticity of a linear chain for unbranched trees with quenched $L=l_K N$.
The first term of Eq.~(\ref{eq:fGutinPrefactors}) is the usual elastic energy contribution for stretching a polymer of linear contour length $L$ at its ends~\cite{GutinGrosberg93}.
The second term penalises deviations from the ideal branching statistics, which lead to longer paths and hence spatially more extended trees.
More formally, it is calculated from the partition function of an ideal branched polymer of $N$ bonds with $L$ bonds between two arbitrary fixed ends~\cite{DeGennes1968,GutinGrosberg93,GrosbergNechaev2015}.
For trees with quenched connectivity, Eq.~(\ref{eq:fGutinPrefactors}) has to be evaluated for the given mean path length $L$ and then minimised with respect to $R$.
For trees with annealed connectivity, Eq.~(\ref{eq:fGutinPrefactors}) needs to be minimised with respect to both, $R$ and $L$.

In this form, the theory predicts values for the exponents $\nu$, $\rho$, and $\nu_{path}$ for a wide range of tree systems~\cite{Rosa2016a,Rosa2016b} as a function of the embedding dimension, $d$, as well as relations between these exponents.
For example, optimising $L$ for annealed trees for a {\em given} asymptotic, $R \sim N^\nu$, yields
\begin{eqnarray}
\label{eq:rho_of_nu}
\rho &=&\frac{1+2\nu}3\\
\label{eq:nupath_of_nu}
\nu_{path}&=& \frac{3\nu}{1+2\nu}
\end{eqnarray}
{\em independently} of the type of volume interactions causing the swelling in the first place.
Plausibly, a fully extended system, $\nu=1$,  is predicted not to branch, $\rho=1$, and to have a fully stretched stem, $\nu_{path}=\nu=1$. For the radius of ideal randomly branched polymers, $\nu=1/4$, one recovers $\rho=1/2$ and Gaussian path statistics, $\nu_{path}=1/2$.

\subsection{Linear chains beyond Flory theory}\label{sec:BeyondFlorySAW}
There is more to (linear) polymers than can be described by Flory theory.
The number of self-avoiding walks is given by~\cite{Guttmann1987,DesCloizeauxBook}
\begin{equation}\label{eq:gamma_SAW}
{\mathcal Z}_{SAW}(N) \sim \mu^N N^{\gamma-1}
\end{equation}
with a universal exponent $\gamma$ and a non-universal constant $\mu$ characteristic of the employed lattice.
Flory theory can neither predict the functional form of Eq.~(\ref{eq:gamma_SAW}), nor the numerical value of $\gamma$, nor the related contact probability~\cite{DesCloizeauxBook}
\begin{eqnarray}
p_c &\sim& N^{-\nu(d+\theta)}
\label{eq:theta_SAW}\\
\theta &=& \frac{\gamma-1}{\nu} \, .
\label{eq:theta_gamma_nu}
\end{eqnarray}
Furthermore, Flory theory incorrectly predicts that stretched chains exhibit essentially Gaussian behavior with volume interactions becoming quickly negligible.
Instead, the end-to-end distance distribution of self-avoiding walks is to an excellent approximation~\cite{EveraersJPA1995} given by the Redner-des Cloizeaux (RdC)~\cite{Redner1980,DesCloizeauxBook} distribution
\begin{eqnarray}
p_N(\vec r) &=& \frac1{\rsq{N}^{d/2}} \ q\left(\frac{\vec r }{ \sqrt{\rsq{N}}}\right)\\
\label{eq:RdC}
q(\vec x) &=&  C \, x^\theta\  \exp \left( -(K x)^t \right) 
\end{eqnarray}
with $q(\vec x)$ independent of $N$. For small distances, Eq.~(\ref{eq:RdC}) is dominated by the power law with the exponent $\theta$ given by the contact exponent Eq.~(\ref{eq:theta_SAW}). For large distances, the chain behaves like a string of $N/g$ blobs of size $\xi\sim g^\nu$. For a given extension $r/\xi$, the free energy of $k_BT$ per blob implies that the exponents $t$ is given by~\cite{FisherSAWShape1966,PincusBlob1976} 
\begin{eqnarray}
t &=& \frac{1}{1-\nu}\ .
\label{eq:t_SAW}
\end{eqnarray}
Interestingly, knowledge of the two exponents is sufficient to reconstruct the entire distribution function, because the constants $C$ and $K$ are determined by the conditions (1) that the distribution is normalized ($\int q(\vec x) d\vec x \equiv 1 $) and (2) that the second moment was chosen as the scaling length ($\int |x|^2 q(\vec x) d\vec x \equiv 1 $):
\begin{eqnarray}
C &=& t \, \frac{\Gamma(1+\frac d2)\Gamma^{\frac{d+\theta}2}(\frac{2+d+\theta}t)}{d\,\pi^{d/2}\,\Gamma^{\frac{2+d+\theta}2}(\frac{d+\theta}t)}
\label{eq:RdC_C}\\
K^2 &=& \frac{\Gamma(\frac{2+d+\theta}t)}{\Gamma(\frac{d+\theta}t)} \ .
\label{eq:RdC_K}
\end{eqnarray}

\subsection{Lee-Yang edge singularity}
Lattice trees are believed to fall into the same universality class as lattice animals~\cite{LubenskyIsaacson1979,SeitzKlein1981,DuarteRuskin1981} with their number scaling similarly to Eq.~(\ref{eq:gamma_SAW}):
\begin{equation}\label{eq:gamma_TREE}
{\mathcal Z}_{SAT}(N) \sim \mu^N N^{\gamma-1}
\end{equation}
In particular, the tree and animal critical exponents in $d$-dimensions are related to the Lee-Yang edge singularity~\cite{ParisiSourlasPRL1981,FisherPRL1978,KurtzeFisherPRB1979,BovierFroelichGlaus1984} of the Ising model in an imaginary magnetic field in $(d-2)$-dimensions, suggesting a relation between the entropy and the size
\begin{equation}\label{eq:LeeYang}
\gamma(d) = -\nu(d)(d-2) \, .
\end{equation}
Interestingly, this relation suggests that it is possible to estimate the number of self-avoiding trees using Flory theory.

\section{Model and methods}\label{sec:modmethods}

In this section, we account very briefly for the algorithms used for generating equilibrated configurations of trees (Sec.~\ref{sec:TreeEnsembles})
and the numerical schemes employed for their analysis (Sec.~\ref{sec:TreesAnalysis}).
The reader interested in more technical details may look into our former works~\cite{Rosa2016a,Rosa2016b}.
A longer discussion is dedicated to how we extracted and extrapolated scaling exponents for RdC functions and contacts (Sec.~\ref{sec:FiniteSize}).
Quantitative details as well as tabulated values for single-tree statistics are presented in the Supplemental Material.

\subsection{Generation of equilibrated tree configurations for different ensembles}\label{sec:TreeEnsembles}

To simulate randomly branching polymers with {\it annealed} connectivity,
we employ a slightly modified version of the ``amoeba'' Monte-Carlo algorithm~\cite{SeitzKlein1981} for trees on the cubic lattice with periodic boundary conditions.
In the model, connected nodes occupy adjacent lattice sites.
As there is no bending energy term, the lattice constant equals the Kuhn length, $l_K$, of linear paths across ideal trees.
The functionality of nodes is restricted to the values $f=1$ (a leaf or branch tip), $f=2$ (linear chain section), and $f=3$ (branch point).
We have studied ideal non-interacting trees, $3d$ self-avoiding trees, and $2d$ and $3d$ melts of trees.

In addition, we have studied randomly branched trees with {\it quenched ideal} connectivity.
For this ensemble, we have resorted to an equivalent off-lattice bead-spring model and studied it via Molecular Dynamics simulations. 
In this model, trees are represented by the same number of degrees of freedom, with bonds described as harmonic springs of average length equal to $l_K$.
This allows a one-to-one mapping to and from the lattice model.
Furthermore, beads interact with a repulsive soft potential whose strength is tuned in such a way, that the gyration radii of self-avoiding lattice trees remain invariant under the switch to the off-lattice model, if their quenched connectivities are drawn from the ensemble of self-avoiding on-lattice trees with annealed connectivity.

For all our ensembles, the tree sizes are
$3 \leq N \leq 1800$ for ideal and self-avoiding trees~\cite{Rosa2016a}
and
$3 \leq N \leq 900$ for melts of trees~\cite{Rosa2016b}.

\subsection{Analysis of tree connectivity}\label{sec:TreesAnalysis}

We have analysed tree connectivities using a variant of the ``burning'' algorithm for percolation clusters~\cite{StanleyJPhysA1984,StaufferAharonyBook}.
The algorithm works iteratively:
each step consists in removing from the list of all nodes the ones with functionality $=1$
and updating the functionalities and the indices of the remaining ones accordingly.
The algorithm stops when only one node (the ``center'' of the tree) remains in the list.
In this way, by keeping track of the nodes which have been removed it is possible to obtain information about the mass and shape of branches.
The algorithm can be then generalized to detect the minimal path length $l_{i,j}$ between any pair of nodes $i$ and $j$:
it is in fact sufficient that both nodes ``survive'' the burning process.

\subsection{Finite-size effects}\label{sec:FiniteSize}
As discussed in detail in our former works~\cite{Rosa2016a,Rosa2016b},
extrapolation to the large-$N$ limit of scaling exponents is a delicate issue.
In general, in fact, our data are affected by finite-size effects and extracted exponents are either
(i) effective (crossover) exponents valid for the particular systems and system sizes we have studied or
(ii) {\em estimates} of true, asymptotic exponents, which suffer from uncertainties related to the extrapolation to the asymptotic limit. 

{\it Extrapolating scaling exponents of distribution functions} --
Pairs of exponents
($\theta_l, t_l$),
($\theta_{path}, t_{path}$),
and
($\theta_{tree}, t_{tree}$)
are obtained by best fits of data for distribution functions $p_N(l)$, $p_N({\vec r} | l)$ and $p_N({\vec r})$
to the corresponding $2$-parameter Redner-des Cloizeaux functions (Eqs.~(\ref{eq:q_RdC_l}), (\ref{eq:q_RdC_path}) and~(\ref{eq:q_RdC_tree}), respectively).
As shown in Tables~SI-III 
the values obtained from these fits display non negligible finite-size effects.
Then, the search for extrapolated values has required two separate strategies.

For exponents ($\theta_l, t_l$) and ($\theta_{tree}, t_{tree}$),
we follow a procedure similar to the one outlined first in~\cite{MadrasJPhysA1992} and adopted later by us in~\cite{Rosa2016a,Rosa2016b}.
It combines together the two following extrapolation schemes:
\begin{enumerate}
\item
A fit of the data for $\theta_l$ and $\theta_{tree}$ 
to the following 3-parameter fit functions:
\begin{eqnarray}
\log \theta_l
& = & a + b N^{-\Delta_0} - b  (\Delta-\Delta_0) N^{-\Delta_0 } \log N \nonumber\\
& \equiv & a + b e^{-\Delta_0 \log N} - b  (\Delta-\Delta_0) e^{-\Delta_0 \log N} \log N \nonumber\\
& & \label{eq:ExtrapolateThetalFuncts}
\end{eqnarray}
and
\begin{eqnarray}
\theta_{tree}
& = & a + b N^{-\Delta_0} - b  (\Delta-\Delta_0) N^{-\Delta_0 } \log N \nonumber\\
& \equiv & a + b e^{-\Delta_0 \log N} - b  (\Delta-\Delta_0) e^{-\Delta_0 \log N} \log N \nonumber\\
& & \label{eq:ExtrapolateThetaTreeFuncts}
\end{eqnarray}
and analogous expressions for $t_l$ and $t_{tree}$.
Eqs.~(\ref{eq:ExtrapolateThetalFuncts}) and~(\ref{eq:ExtrapolateThetaTreeFuncts}) correspond to a self-consistent linearisation of the 3 parameter fit
$\theta_{l,tree} = a + b \frac{1}{N^{\Delta}}$ around $\Delta = \Delta_0$.
We have carried out a one-dimensional search for
the value of $\Delta_0$ for which the fits yield vanishing $N^{-\Delta_0} \log N$ term.
Note that we have analyzed data for $\theta_l$ (and $t_l$) in Eq.~(\ref{eq:ExtrapolateThetalFuncts}) in the form ``$\log \theta_l$ {\it vs.} $\log N$'' (log-log),
while for $\theta_{tree}$ (and $t_{tree}$) we have used in Eq.~(\ref{eq:ExtrapolateThetaTreeFuncts}) data in a log-linear representation, ``$\theta_{tree}$ {\it vs.} $\log N$.'' 
These two different functional forms have been found to produce the best (statistical significant) fits. 
\item
In the second method we fixed $\Delta =1$,
and we calculated the corresponding $2$-parameter best fits to the same data. 
\end{enumerate}
Results from the two fits are summarized separately in
Tables~SI 
and~SIII 
and their averages taken for our final estimates of scaling exponents (see the corresponding boldfaced numbers).
In the tables, we have also reported which ranges of $N$ have been considered for the best fits to the data.
Unfortunately, for $\theta_{path}$ and $t_{path}$ an analogous scheme can not be applied because we have data only for very limited ranges of $l$.
Then, our best estimates come from simply averaging single values together (see boldfaced numbers in Table~SII). 

{\it Extrapolating scaling exponents of tree contacts} --
Tabulated values for intra-chain contacts, $\langle N_c(N) \rangle \sim N^{\gamma_c}$, and
inter-chain contacts in tree melts, $\langle N_c^{inter}(N) \rangle \sim N^{\beta}$, are given in Table~SIV. 
Corresponding extrapolated values of critical exponents $\gamma_c$ and $\beta$ were obtained by the same methodology reported in our articles~\cite{Rosa2016a,Rosa2016b}.
For brevity, it was summarized in the caption of Table~SIV. 
Otherwise, the interested reader can look into the above mentioned works for more details.

In all cases, the quality of the fits is estimated by standard statistical analysis~\cite{NumericalRecipes}:
normalized $\chi$-square test ${\tilde \chi}^2 \equiv \frac{\chi^2}{D-f}$,
where $D - f$ is the difference between the number of data points, $D$, and the number of fit parameters, $f$.
When ${\tilde \chi}^2 \approx 1$ the fit is deemed to be reliable.
The corresponding $\mathcal Q(D-f, \chi^2)$-values provide a quantitative indicator for the likelihood that $\chi^2$ should exceed the observed value, if the model were correct~\cite{NumericalRecipes}.
The results of all fits (tables~SI-IV) 
are reported together with the corresponding errors, ${\tilde \chi}^2$ and $\mathcal Q$ values. 
Unless otherwise said, all error bars for the estimated asymptotic values (boldfaced numbers in Tables~SI-IV) 
are written in the form $\pm$(statistical error)$\pm$(systematic error),
where the ``statistical error'' is the largest value obtained from the different fits~\cite{MadrasJPhysA1992} and the ``systematic error'' is the spread between the single estimates.
For brevity, these are combined together into one single error bar in Table~\ref{tab:ExpSummary} as:
$\sqrt{(\mbox{statistical error})^2 + (\mbox{systematic error})^2}$.

\section{Results and Discussion}\label{sec:results}
We begin with distribution functions for observables characterizing the tree connectivity: the distribution of branch weights (Sec.~\ref{sec:BranchWeightStatistics}) and path lengths (Sec.~\ref{sec:ppStat}).
Then, we turn to the conformational statistics of linear path on the tree (Sec.~\ref{sec:ConfStatPaths}) and the distribution of internal distances  (Sec.~\ref{sec:TreeStat}).

\subsection{Branch weight statistics and generalised Kramers relation}\label{sec:BranchWeightStatistics}

\begin{figure}
\includegraphics[width=0.49\textwidth]{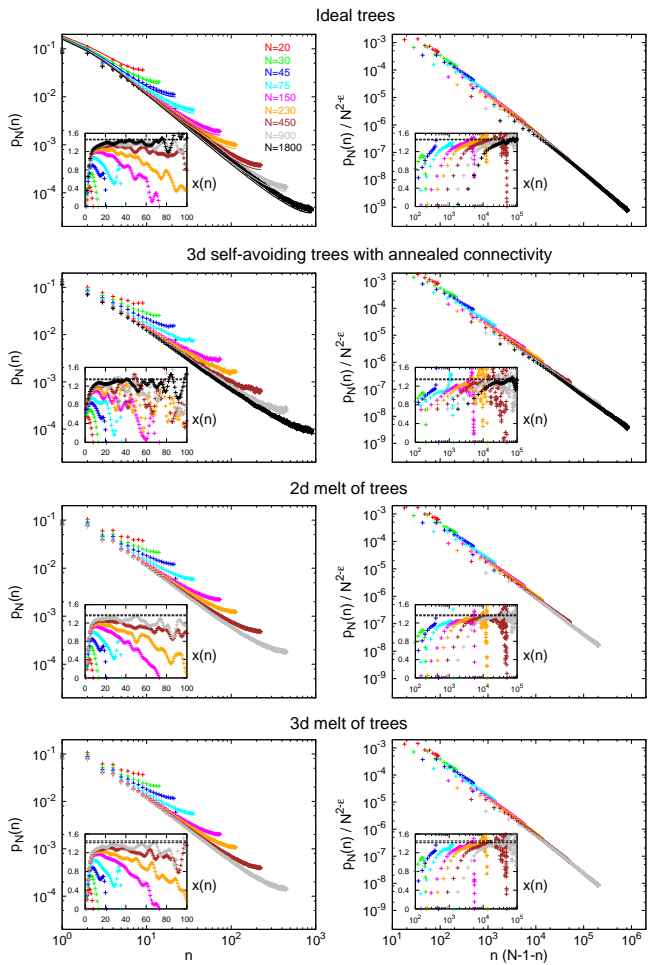}
\caption{
\label{fig:pcutBranch}
(Left-hand panels)
Probability distribution functions for branch weight $n$, $p_N(n)$, in trees of total mass $N$.
(Right-hand panels)
$N^{2-\epsilon}$-rescaled distributions as functions of $n(N-1-n)$.
In the top panel on the left, solid lines correspond to the exact analytical solution by Daoud and Joanny, Eq.~(\ref{eq:DaoudJoanny_SplitProb}).
Left-hand and right-hand insets:
Corresponding differential fractal exponents
$x = x(n) \equiv - \frac{\log \, p_N(n+1) \, / \, p_N(n)}{\log \, (n+1) \, / \, n}$
and
$x = x(n) \equiv - \frac{\log \, p_N \left((n+1)(N-2-n)\right) \, / \, p_N \left(n(N-1-n)\right)}{\log \, \left((n+1)(N-2-n)\right) \, / \, \left(n(N-1-n)\right)}$
(data were smoothed for better visualization)
are compared to the asymptotic scaling relation $\lim_{n \rightarrow \infty} x(n) = 2-\epsilon$
(the regions within dashed lines,
corresponding to the estimated values of ``$\epsilon \pm \mbox{(error bars)}$'' as reported in Refs.~\cite{Rosa2016a,Rosa2016b}).
}
\end{figure}

In Figure~\ref{fig:pcutBranch} we show the distribution, $p_N(n)$, of the weight, $n$, of the branches generated by cutting randomly selected bonds in trees of size $N$.
Obviously, it is possible to cut larger branches from larger trees, but independently of $N$ the vast majority of the branches is small.
This follows immediately from the fact that for all our systems a large fraction ($\approx 40\%$ for ideal and melt of trees and $\approx 27\%$ for annealed self-avoiding trees~\cite{Rosa2016a,Rosa2016b}) of the nodes is one-functional:
cutting the bonds joining them to the tree generates branches of weight $n=0$.

To gain some intuition for the form of these distributions, it is useful to reconsider the case of ideal trees (Section~\ref{sec:intro}, Eqs.~(\ref{eq:DJpartfunct})-(\ref{eq:DaoudJoanny_SplitProb})), 
where ${\mathcal Z}_n \sim e^{2\lambda n} / (\lambda n)^{3/2}$.
In that case, $p_N(n) \sim \frac{N^{3/2}}{n^{3/2}(N-n-1)^{3/2}}$, which simplifies to $p_N(n) \sim n^{-3/2}$ for small $1< n \ll N$.
As expected, by plotting data for $p_N(n)$ in log-log plots as a function of $n$ (l.h. column of Fig.~\ref{fig:pcutBranch}) 
we find good agreement with the expected power law for small $n$, while plotting data as a function of $n(N-1-n)$ (r.h.s) produces nearly perfect power law behavior over the entire range $1\le n\le N/2$.

Interestingly, we find similar power law behaviour for interacting trees, too.
We therefore tentatively generalize Eq.~(\ref{eq:DJpartfunct}) for the tree partition function to 
\begin{equation}
{\mathcal Z}_n \sim c^n n^{-x} \ .
\end{equation}
In fact, this form is compatible with the Kramers theorem, Eq.~(\ref{eq:Kramers_SplitProb}),
since $ \sum_{n=0}^{N-1} {\mathcal Z}_n \, {\mathcal Z}_{N-1-n}= c^N \sum_{n=0}^{N-1} n^{-x} (N-1-n)^{-x} \sim c^N  N^{-x} =  {\mathcal Z}_N$.
The resulting average branch weight of $\langle N_{br}(N) \rangle \sim N^{2-x}$ implies $x=2-\epsilon$.
As shown in the corresponding insets, the relation $x=2-\epsilon$ is well satisfied with values for $\epsilon$ taken from Refs.~\cite{Rosa2016a,Rosa2016b}.
The numerical prefactor $c$ is related to the asymptotic branching probability (see Fig.~4 in Ref.~\cite{Rosa2016a} and Fig.~3 in Ref.~\cite{Rosa2016b}),
since in the limit $N\rightarrow\infty$, 
$\langle n_3\rangle/N = \langle n_1\rangle/N = 2p_N(n=0)$~\cite{Rosa2016a}, and 
$p_N(n=0)={\mathcal Z}_{N-1}/{\mathcal Z}_N=c^{-1} \left(\frac{N-1}N \right)^{\epsilon-2} \simeq c^{-1}$.

\subsection{Path length statistics for trees}\label{sec:ppStat}

\begin{figure}
\vspace{2mm}
\includegraphics[width=0.49\textwidth]{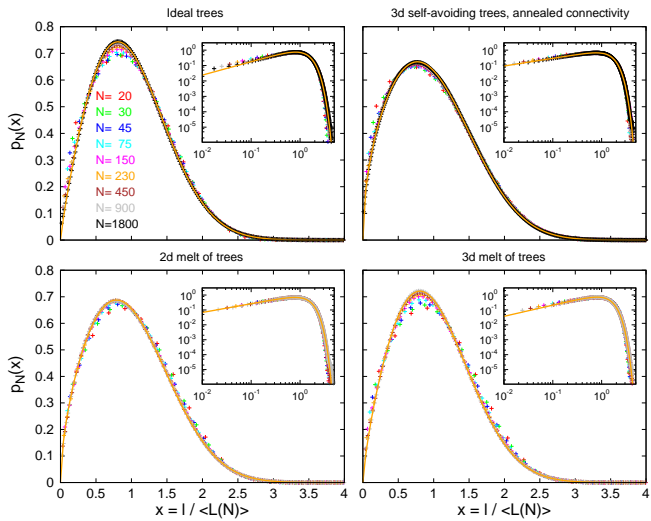}
\caption{
\label{fig:PathL_PDF}
Probability distribution functions, $p_N(l)$, for linear paths of length $l$ and average size $\langle L(N) \rangle$ in trees of total mass $N$.
Orange solid lines correspond to the one-dimensional Redner-des Cloizeaux function, Eq.~(\ref{eq:q_RdC_l}),
with parameters $\theta_l$ and $t_l$ obtained from the best fits to data for
$N=1800$ (ideal and self-avoiding trees) and
$N=900$ ($2d$ and $3d$ melt of trees),
see Table~SI 
for detailed values of fit parameters.
Inset: Same plots in log-log scale.
}
\end{figure}

As illustrated in Fig.~\ref{fig:PathL_PDF}, the measured path length distribution functions, $p_N(l)$,
fall onto universal master curves, when plotted as a function of the rescaled path length
$x = l / \langle L(N) \rangle$:
\begin{equation}\label{eq:pl}
p_N(l) = \frac1{\langle L(N) \rangle}\  q \left(\frac{l}{\langle L(N) \rangle}\right) \, .
\end{equation}
These master curves are well described by the one-dimensional Redner-des Cloizeaux (RdC) form (orange lines in Fig.~\ref{fig:PathL_PDF}):
\begin{equation}
q(x) = C_l \, x^{\theta_l}\  \exp \left( -(K_l x)^{t_l} \right) \, .
\label{eq:q_RdC_l}
\end{equation}
The constants
\begin{eqnarray}
C_l &=& t_l \, \frac{\Gamma^{\theta_l+1}((\theta_l+2)/t_l)}{\Gamma^{\theta_l+2}((\theta_l+1)/t_l)}
\label{eq:RdC_C_l}\\
K_l &=& \frac{\Gamma((\theta_l+2)/t_l)}{\Gamma((\theta_l+1)/t_l)}
\label{eq:RdC_K_l}
\end{eqnarray}
follow from the conditions that $p_N(l)$ is normalized to $1$ and that the {\it first} moment, $\langle L(N) \rangle$, is the only relevant scaling variable.
Estimated values for $(\theta_l, t_l)$ obtained from best fits of Eq.~(\ref{eq:q_RdC_l}) to data for specific values of $N$
and extrapolated values to large $N$ are summarized in Table~SI. 

Interestingly, we can give a physical interpretation of the observed (effective) exponents.
For small path lengths, results are not affected by the total tree size.
We thus expect to find $n(l_{max})\sim l_{max}^{1/\rho}$ segments at a contour distance $l \le l_{max}$ from any node.
Since $p(l_{max}) \sim d n(l_{max}) / d l_{max}$, this suggests the scaling relationship
\begin{equation}\label{eq:theta_l}
\theta_l = \frac{1}{\rho}-1 \, .
\end{equation}
To estimate the probability for observing very long paths, it is tempting to adjust the Pincus-blob argument~\cite{PincusBlob1976} cited in the Section~\ref{sec:BeyondFlorySAW}.
A stretched tree should behave like a string of $N/g$ unperturbed trees of size $\xi\sim g^\rho$, suggesting that
\begin{equation}\label{eq:tl}
t_l = \frac{1}{1-\rho}\ .
\end{equation}
The argument works well (see bottom panel of Table~\ref{tab:ExpSummary}, columns (a) and (b)) when
comparing the asymptotic values for $t_l$ in different ensembles to the corresponding numerical values for $\rho$ taken from~\cite{Rosa2016a,Rosa2016b}.
Interestingly, being only functions of $\rho$, $\theta_l$ and $t_l$ can also be explicitly calculated in terms of results for $\rho$ from the Flory theory~\cite{Rosa2016a,Rosa2016b},
see top panel of Table~\ref{tab:ExpSummary}.
Interestingly, the Flory theory gives a remarkable accurate prediction for most of the cases discussed.

\subsection{Conformational statistics of linear paths}\label{sec:ConfStatPaths}

\begin{figure}
\includegraphics[width=0.5\textwidth]{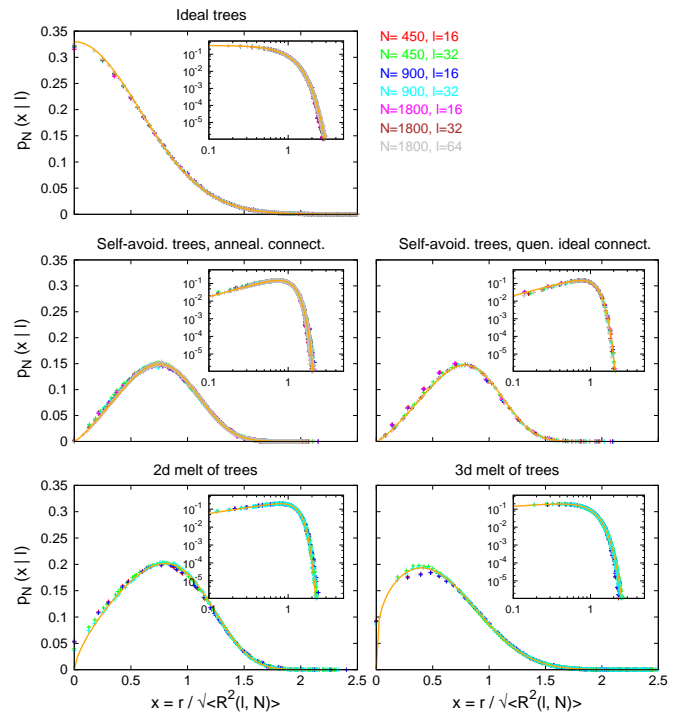}
\caption{
\label{fig:Path_E2EDistance_Distribution}
Probability distribution functions, $p_N(r|l)$, of spatial distances, $r$, for linear paths of contour length $l$.
Orange solid lines 
correspond to theoretical predictions:
for ideal trees data match the Gaussian distribution,
while melt of trees and self-avoiding trees are well described by the Redner-des Cloizeaux function, Eq.~(\ref{eq:q_RdC_path}),
with parameters $\theta_{path}$ and $t_{path}$ obtained from the best fits to data
for $N=1800$ and $l=64$ (self-avoiding trees)
and $N=900$ and $l=32$ ($2d$ and $3d$ melt of trees),
see Table~SII 
for detailed values of fit parameters.
Insets: Same plots in log-log representation.
}
\end{figure}

Fig.~\ref{fig:Path_E2EDistance_Distribution} shows measured end-to-end vector distributions, $p_N(\vec r|l)$, for paths of length $l$ on trees of mass $N$. 
The data superimpose, when expressed as functions of the scaled distances, $x = \left| \vec r \right| / \sqrt{\rsq{l,N}}$:
\begin{equation}
p_N(\vec r|l) = \frac1{\rsq{l,N}^{3/2}}\  q\left(\frac{\left| \vec r \right|  }{ \sqrt{\rsq{l,N}}}\right) \, .
\label{eq:pr_of_l}
\end{equation}
Moreover, they are in excellent agreement with the RdC distribution (Eqs.~(\ref{eq:RdC}), (\ref{eq:RdC_C}) and (\ref{eq:RdC_K}) and orange lines in Fig.~\ref{fig:Path_E2EDistance_Distribution}):
\begin{equation}
q(x) = C \, x^{\theta_{path}}\  \exp \left( -(K x)^{t_{path}} \right) \, .
\label{eq:q_RdC_path}
\end{equation}
The shape of the rescaled distributions, and hence the characteristic exponents $\theta_{path}$ and $t_{path}$, depend on the universality class.
Not surprisingly, paths on ideal trees are well described by the Gaussian distribution, i.e. $\theta_{path}=0$ and $t_{path}=2$.
Fitted values for the other cases are listed in Table~SII. 
Given the limited range of available path lengths $l$, we have found no meaningful way to estimate asymptotic values.
We have then simply taken the average of the available fitted values (see boldfaced numbers in Table~SII). 
Again, the observed values can be given a physical interpretation.

The exponent $\theta_{path}$ describes the reduction of the contact probability, Eq.~(\ref{eq:theta_SAW}), relative to a na{\"i}ve Gaussian estimate.
Importantly, it is a genuinely novel exponent, {\it i.e.} it is {\it independent} from all other exponents discussed in Refs.~\cite{Rosa2016a,Rosa2016b} and in this work.
In the case of self-avoiding walks, $\theta$ is related to the entropy exponent $\gamma$, Eq.~(\ref{eq:theta_gamma_nu})~\cite{Guttmann1987,DesCloizeauxBook}.
Interestingly, Grosberg and colleagues argued~\cite{GutinGrosberg93} that the identical Flory predictions of $\nu=3/(d+2)$ for self-avoiding walks and for the path statistics in melts of annealed lattice trees suggests a deeper analogy between the two problems~\cite{GrosbergSoftMatter2014}.
Using $\gamma_{2d} \approx 1.344$ and $\gamma_{3d} \approx 1.162$ for two and three-dimensional self-avoiding walks~\cite{Guttmann1987} and Eq.~(\ref{eq:theta_gamma_nu}) suggests $\theta_{path, 2d} \approx 0.459$ and $\theta_{path, 3d} \approx 0.276$.
In particular, the $3d$ value is in very good agreement with our finding $\theta_{3d} = 0.28 \pm 0.02$ (Tables~\ref{tab:ExpSummary} and~SII), 
while the $2d$ value appears smaller than the reported $\theta_{2d} = 0.63 \pm 0.04$.
Notice though, that these exponents were measured for path length $l={\cal O}(50)$ and then finite-size effects may likely induce a bias on the final result.

The exponent $t_{path}$ controls the non-linear path elasticity at large elongations.
The measured effective values can be compared to the Fisher-Pincus relation, Eq.~(\ref{eq:t_SAW}),
for self-avoiding walks~\cite{FisherSAWShape1966,PincusBlob1976}
\begin{equation}\label{eq:tPath}
t_{path} = \frac{1}{1-\nu_{path}} \, ,
\end{equation}
where specific values for $\nu_{path}$ are taken from Refs.~\cite{Rosa2016a,Rosa2016b},
see bottom panel of Table~\ref{tab:ExpSummary} (columns (a) and (b)).
In general, agreement is overall good. The only exception is for $3d$ self-avoiding trees with quenched ideal statistics,
which, again, may be ascribed to the limited range of path lengths of our simulated trees.
As for $\theta_l$ and $t_l$ and being a function of $\nu_{path}$ only,
specific values for $t_{path}$ can be also obtained by using the Flory results~\cite{Rosa2016a,Rosa2016b} for $\nu_{path}$ (top panel of Table~\ref{tab:ExpSummary}):
again, for most of the cases, there exist fair agreement with numerical predictions.

\subsection{Conformational statistics of trees}\label{sec:TreeStat}

\begin{figure}
\includegraphics[width=0.5\textwidth]{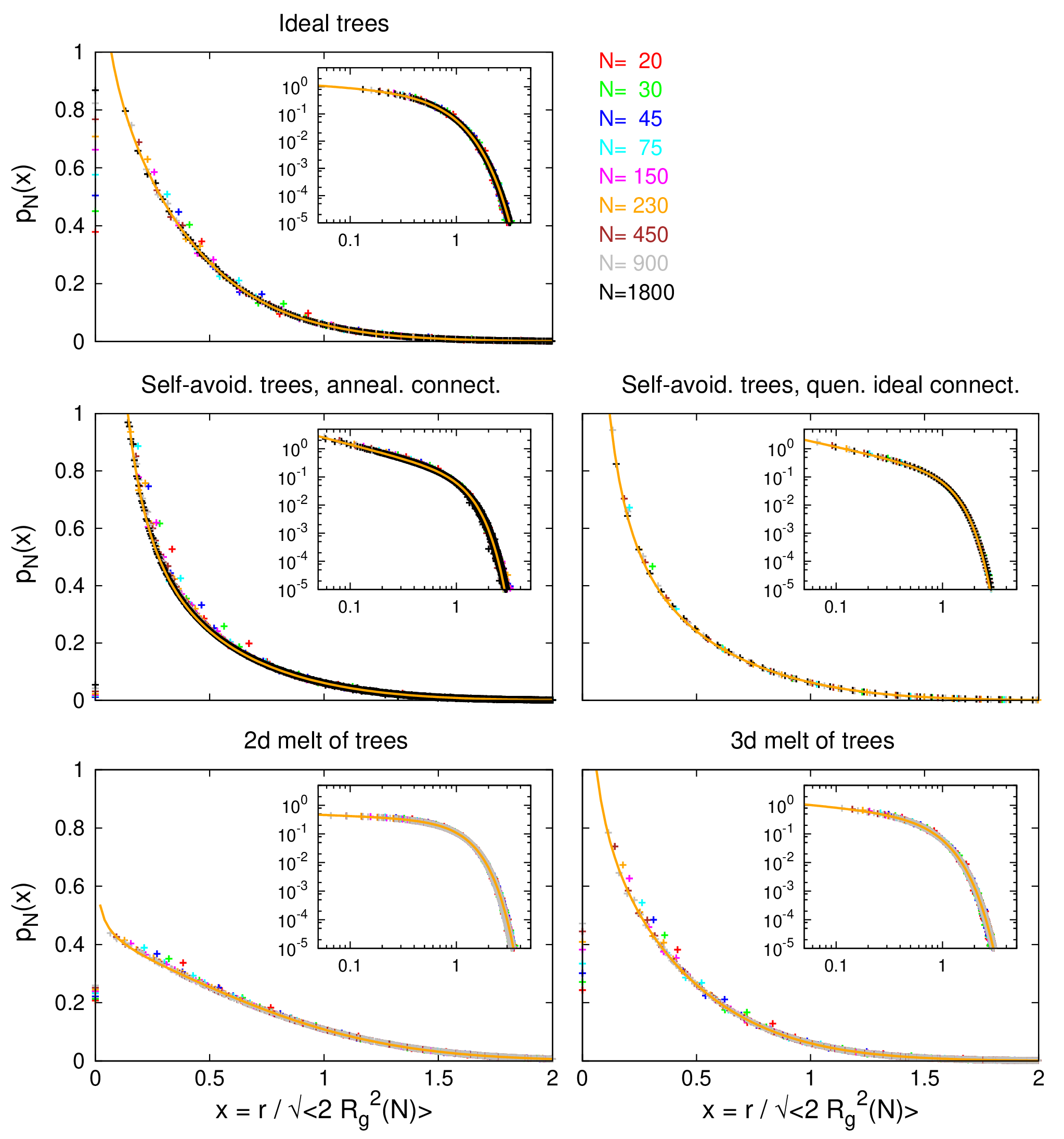}
\caption{
\label{fig:MonMonDistsPDF_CoMDistsPDF}
Probability distribution functions, $p_N(r)$, of spatial distances, $r$, between pairs of nodes.
Orange solid lines correspond to the Redner-des Cloizeaux distribution function, Eq.~(\ref{eq:q_RdC_tree}),
with parameters $\theta_{tree}$ and $t_{tree}$ obtained from the best fits to data with $N=1800$ (ideal and self-avoiding trees) and $N=900$ ($2d$ and $3d$ melt of trees),
see Table~SIII 
for detailed values of fit parameters.
Insets: same plots in log-log scale.
}
\end{figure}

Fig.~\ref{fig:MonMonDistsPDF_CoMDistsPDF} (left panels) shows distributions $p_N(\vec r)$ of vectors $\vec r$ connecting all tree nodes. 
The data superimpose, when expressed as functions of the scaled distances, $x = \left| \vec r \right| / \sqrt{2 \langle R_g^2(N) \rangle}$:
\begin{equation}\label{eq:pr}
p_N(\vec r) = \frac1{\left(2 \langle R_g^2(N) \right)^{3/2}}\  q\left(\frac{\left| \vec r \right|  }{ \sqrt{2 \langle R_g^2(N) \rangle}}\right) \, .
\end{equation}
Again, the distributions are in excellent agreement with the RdC form (Eqs.~(\ref{eq:RdC}), (\ref{eq:RdC_C}) and (\ref{eq:RdC_K} and orange lines in Fig.~\ref{fig:MonMonDistsPDF_CoMDistsPDF}):
\begin{equation}\label{eq:q_RdC_tree}
q(x) =  C \, x^{\theta_{tree}}\  \exp \left( -(K x)^{t_{tree}} \right) \, .
\end{equation}
The extracted exponents and corresponding extrapolations to $N\rightarrow \infty$ are listed in Table~SIII. 

In the following, we relate the characteristic exponents $\theta_{tree}$ and $t_{tree}$ to our previous results by using Eqs.~(\ref{eq:q_RdC_path}) and (\ref{eq:q_RdC_l}) together with the convolution identity 
\begin{equation}\label{eq:p_convolution}
p_N(r) = \int_0^{\infty} p_N(r|l) \, p_N(l) \, d l\ \ ,
\end{equation}
which states that the local density can be calculated by adding up the contributions from paths of all possible length, $1\le l\le N$. 
The behavior of $p_N(r)$ for large distances, $r > R_g$, can be estimated from the contour distance $l^\ast(r)$, which makes the dominant contribution to particle pairs found at the spatial distance $r$.
Combining the arguments of the compressed exponentials in Eqs.~(\ref{eq:q_RdC_l}) and~(\ref{eq:q_RdC_path}), this requires the minimization of
$\left( \frac{l}{\langle L(N) \rangle} \right)^{t_l} + \left( \frac{r}{\sqrt{\rsq{l}}} \right)^{t_{path}}$
and yields
\begin{equation}\label{eq:tTree}
t_{tree} = \frac{t_l \, t_{path}} {t_l + t_{path} \, \nu_{path}} = \frac{1}{1-\nu} \, .
\end{equation}
Results from computer simulations for asymptotic exponents $t_{tree}$ and $\nu$~\cite{Rosa2016a,Rosa2016b} support well this relation
(see bottom panel of Table~\ref{tab:ExpSummary}, columns (a) and (b)).

In the limit of small distances, $r<R_g$, there are two possibilities. If there is no power law divergence in the small $l$ limit,  the integral is dominated by contributions from long paths with $\rsq{l}\gg r^2$, allowing to set the exponential term in Eq.~(\ref{eq:q_RdC_path}) equal to one.
The only $r$-dependence comes through the explicit $r^{\theta_{path}}$ term and hence
\begin{equation}
\theta_{tree} =  \theta_{path} \ \ \ \ \mbox{if\ \ \ } \theta_{path}< \frac1\nu-d
\end{equation}
In the opposite limit, short paths dominate.
The apparent divergence of the integrand in the limit $l\rightarrow0$ is removed by the exponential tail of Eq.~(\ref{eq:q_RdC_path}):
paths with vanishing contour lengths up to $l\sim r^{1/\nu_{path}}$ do not contribute to the monomer density for finite spatial distances $r$.
For longer paths, we can set the exponential to one. In this case
\begin{equation}
\theta_{tree} =  \frac{1}{\nu} - d \ \ \ \ \mbox{if\ \ \ }  \frac1\nu-d<\theta_{path}
\end{equation}
Summarizing
\begin{eqnarray}\label{eq:ThetaTree}
\theta_{tree} &=& \min( \theta_{path} , \frac{1}{\nu} - d ) \\
&=& \left\{\begin{array}{ll}
		0&\mbox{for ideal trees in $d\le4$}\\
		\frac{1}{\nu} - d &\mbox{else}
        \end{array} \right.
\, .
\end{eqnarray}
since $\theta_{path} \equiv0$ for ideal trees and since we expect $\theta_{path}>0$ and $\nu\ge1/d$ for interacting trees.
Table~\ref{tab:ExpSummary} compares the asymptotic results to theoretical predictions.
Again, the general agreement is fairly good. 
Finally, as for $\theta_l$, $t_l$ and $t_{path}$, specific values for $\theta_{tree}$ and $t_{tree}$
can be calculated by resorting to the Flory results for $\nu$ (top panel of Table~\ref{tab:ExpSummary}).
Once again, the predictions prove to be remarkably accurate.

To conclude the section, in Fig.~S1 
in Supplemental Material we focus on distribution functions of monomers around the tree center of mass.
Clearly, for the central node this distribution is Gaussian.
On the other hand, distant nodes can not distinguish between the central node and the tree center of mass and their positions hence follow again a RdC distribution.
Averaged over node identities, the monomer distribution around the tree center of mass seems to become Gaussian for self-avoiding trees, but {\em not} for ideal trees or trees in melt.
Interestingly, the latter effect was already noted a long time ago for {\it ideal} linear chains~\cite{DebyeBuecheJCP1952}.

\subsection{Self-contacts}\label{sec:SelfContacts}

\begin{figure*}
\includegraphics[width=\textwidth]{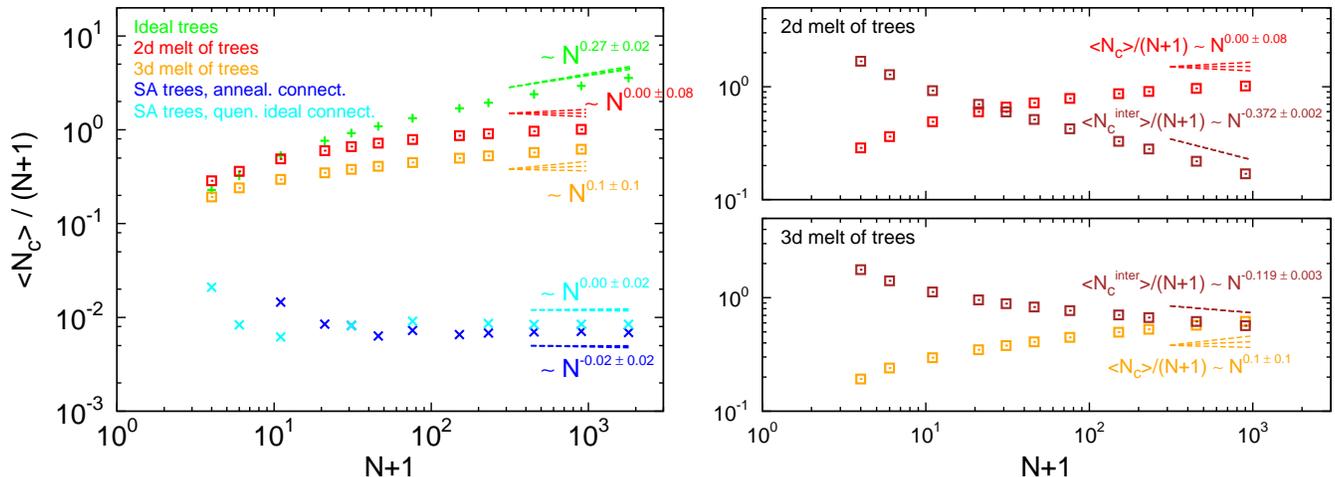}
\caption{
\label{fig:Contacts}
(Left)
Average number of {\it intra}-chain contacts per node, $\langle N_c(N) \rangle / (N+1) \sim N^{\gamma_c-1}$, in the different ensembles.
(Right)
Comparison to the average number of {\it inter}-chain contacts per node, $\langle N_c^{inter} \rangle  / (N+1) \sim N^{\beta-1}$, in $2d$ and $3d$ melts.
Dashed lines mark the ranges of corresponding asymptotic behaviours with scaling exponents $\gamma_c$ and $\beta$, see Table~\ref{tab:ExpSummary}.
}
\end{figure*}

We turn then to the average number of self-contacts per tree,
$\langle N_c (N) \rangle \sim N^{\gamma_c}$ (l.h. panel of Fig.~\ref{fig:Contacts} and Table~SIV). 
Consider an arbitrary pair of monomers. The probability to find them in close contact scales as $N^{-\nu(d+\theta_{tree})}$.
Since there are ${\mathcal O}(N^2)$ different monomer pairs, we have 
\begin{eqnarray}
\gamma_c
&=& 2-\nu(d+\theta_{tree}) \label{eq:gamma_c theta_tree}\\
&=& \left\{\begin{array}{ll} 2-\frac d4 &\mbox{for ideal trees in $d\le4$}\\ 1 &\mbox{else} \end{array} \right.
\, . \nonumber
\end{eqnarray}
This prediction compares extremely well with our numerical estimates for $\gamma_c$, see Table~\ref{tab:ExpSummary}.
Note that the mean-field estimate $\gamma_c=2-d\nu$ holds only for ideal trees in $d\le4$ dimensions.
The melt case is marginal in that we expect $\nu=1/d$ and thus $\gamma_c=1$:
by using the estimated asymptotic values of $\nu$ in $2d$ and $3d$~\cite{Rosa2016b} and $\theta_{tree}$,
the different values for $\gamma_c$ compare well for all studied ensembles (see bottom panel of Table~\ref{tab:ExpSummary}, columns (a) and (b)).
In all other cases, $\gamma_c=1$ independently of $\nu$, $\theta_{path}$ and $\theta_{tree}$, indicating that the local monomer density is finite and independent of tree size. 
This is yet another illustration of the subtle cancellation of errors in Flory arguments, which are built on the mean-field estimates of contact
probabilities~\cite{DeGennesBook}.

For tree melts, we have also considered the average number of contacts between nodes on {\it different} trees, $\langle N_c^{inter}(N) \rangle \sim N^{\beta}$
(r.h. panels of Fig.~\ref{fig:Contacts} and Tables~SIV). 
In the melt, the average number of contacts per node $\frac{\langle N_c(N) \rangle}{N} + \frac{\langle N_c^{inter}(N) \rangle}{N}$ is $N$-independent
(see also Table~SIV), 
hence $\beta \approx 1-\Delta$ where $\Delta$ is the exponent of the power-law correction to the large-$N$ behavior of $\langle N_c (N) \rangle$:
$\langle N_c (N) \rangle \approx a \, N^{\gamma_c}(1 - b \, N^{-\Delta}) = a \, N \, (1 - b \, N^{-\Delta})$ with $a$ and $b$ numerical prefactors.
$\Delta$ can be calculated by considering the two leading terms in Eq.~(\ref{eq:p_convolution})
for small $r$'s after substituting the upper bound of the integral with ${\mathcal O}(N^\rho)$.
Since $\theta_{path} > \frac{1}{\nu}-d$ we get
$p_N(r) \sim \left( \frac{r}{N^\nu} \right)^{1/\nu-d} \left( 1-\mathcal{O} \left( \frac{r}{N^\nu} \right)^{d+\theta_{path}-1/\nu} \right)$.
The average number of self-contacts per tree $\langle N_c(N) \rangle$ is proportional to the integral of the former expression from $0$ to some small cut-off spatial distance, or
$\langle N_c(N) \rangle = a \, N \, (1 - b \, N^{1-\nu(d+\theta_{path})})$. 
Consequently, 
$\Delta = -1+\nu(d+\theta_{path})$ and
\begin{equation}\label{eq:betaDef2}
\beta = 1 - \Delta = 2 - \nu(d+\theta_{path}) \, .
\end{equation}
In particular, Eq.~(\ref{eq:betaDef2}) implies that $\beta < 1$.
By employing the asymptotic values of $\nu$~\cite{Rosa2016b} and $\theta_{path}$, Eq.~(\ref{eq:betaDef2}) shows good agreement with direct estimates of $\beta$ values
(see bottom panel of Table~\ref{tab:ExpSummary}, columns (a) and (b)).

\begin{table*}[h]
\begin{tabular}{|c|c|c|c|c|c|c|}
\multicolumn{7}{c}{Flory theoretical values of critical exponents.}\\
\multicolumn{7}{c}{}\\
\hline
& {\footnotesize Relation to} & {\footnotesize Ideal trees,} & {\footnotesize $2d$ melt of trees,} & {\footnotesize $3d$ melt of trees,} & {\footnotesize $3d$ self-avoiding trees,} & {\footnotesize $3d$ self-avoiding trees,}\\
& {\footnotesize other exponents} & {\footnotesize annealed connect.} & {\footnotesize annealed connect.} & {\footnotesize annealed connect.} & {\footnotesize annealed connect.} & {\footnotesize quenched ideal connect.}\\
\hline
& & & & & & \\
{\footnotesize $\theta_l$} & $=\frac{1}{\rho}-1$ & {\footnotesize $1$} & {\footnotesize $\frac{1}{2}=0.5$} & {\footnotesize $\frac{4}{5}=0.8$} & {\footnotesize $\frac{4}{9} \approx 0.444$} & -- \\
& & & & & & \\
{\footnotesize $t_l$} & $=\frac{1}{1-\rho}$ & {\footnotesize $2$} & {\footnotesize $3$} & {\footnotesize $\frac{9}{4}=2.25$} & {\footnotesize $\frac{13}{4} = 3.25$} & -- \\
& & & & & & \\
{\footnotesize $\theta_{path}$} & -- & {\footnotesize $0$} & {\footnotesize $>0$} & {\footnotesize $>0$} & {\footnotesize $>0$} & {\footnotesize $>0$} \\
& & & & & & \\
{\footnotesize $t_{path}$} & $=\frac{1}{1-\nu_{path}}$ & {\footnotesize $2$} & {\footnotesize $4$} & {\footnotesize $\frac{5}{2} = 2.5$} & {\footnotesize $\frac{9}{2} = 4.5$} & {\footnotesize $\infty$} \\
& & & & & & \\
{\footnotesize $\theta_{tree}$} & {\footnotesize $\min(\theta_{path},\frac{1}{\nu}-d)$} & {\footnotesize $0$} & {\footnotesize $0$} & {\footnotesize $0$} & {\footnotesize $-\frac{8}{7} \approx -1.143$} & {\footnotesize $-1$} \\
& & & & & & \\
{\footnotesize $t_{tree}$} & $=\frac{1}{1-\nu}$ & {\footnotesize $\frac{4}{3} \approx 1.333$} & {\footnotesize $2$} & {\footnotesize $\frac{3}{2}=1.5$} & {\footnotesize $\frac{13}{6} \approx 2.167$} & {\footnotesize $2$} \\
& & & & & & \\
{\footnotesize $\gamma_c$} & {\footnotesize $=2-\nu(d+\theta_{tree})$} & {\footnotesize $\frac{5}{4} = 1.25$} & {\footnotesize $1$} & {\footnotesize $1$} & {\footnotesize $1$} & {\footnotesize $1$} \\
& & & & & & \\
{\footnotesize $\beta$} & {\footnotesize $=2-\nu(d+\theta_{path})$} & -- & {\footnotesize $1-\frac{\theta_{path}}{2} < 1$} & {\footnotesize $1-\frac{\theta_{path}}{3} < 1$} &  -- & -- \\
& & & & & & \\
\hline
\end{tabular}
\begin{tabular}{|c||c|c||c|c||c|c||c|c||c|c|}
\multicolumn{11}{c}{}\\
\multicolumn{11}{c}{}\\
\multicolumn{11}{c}{Numerical values of critical exponents.}\\
\multicolumn{11}{c}{}\\
\hline
& \multicolumn{2}{c||}{\footnotesize Ideal trees,} & \multicolumn{2}{c||}{\footnotesize $2d$ melt of trees,} & \multicolumn{2}{c||}{\footnotesize $3d$ melt of trees,} & \multicolumn{2}{c||}{\footnotesize $3d$ self-avoiding trees,} & \multicolumn{2}{c|}{\footnotesize $3d$ self-avoiding trees,} \\
& \multicolumn{2}{c||}{\footnotesize annealed connect.} & \multicolumn{2}{c||}{\footnotesize annealed connect.} & \multicolumn{2}{c||}{\footnotesize annealed connect.} & \multicolumn{2}{c||}{\footnotesize annealed connect.} & \multicolumn{2}{c|}{\footnotesize quenched ideal connect.} \\
\hline
& {\footnotesize (a)} & {\footnotesize (b)} & {\footnotesize (a)} & {\footnotesize (b)} & {\footnotesize (a)} & {\footnotesize (b)} & {\footnotesize (a)} & {\footnotesize (b)} & {\footnotesize (a)} & {\footnotesize (b)} \\
\hline
& & & & & & & & & & \\
{\footnotesize $\theta_l$} & {\footnotesize $1.1 \pm 0.2$} & {\footnotesize $1.0 \pm 0.2$} & {\footnotesize $0.593 \pm 0.003$} & {\footnotesize $0.63 \pm 0.02$} & {\footnotesize $0.85 \pm 0.07$} & {\footnotesize $0.9 \pm 0.2$} & {\footnotesize $0.53 \pm 0.02$} & {\footnotesize $0.56 \pm 0.05$} & -- & -- \\
& & & & & & & & & & \\
{\footnotesize $t_l$} & {\footnotesize $2.0 \pm 0.1$} & {\footnotesize $2.0 \pm 0.2$} & {\footnotesize $2.35 \pm 0.01$} & {\footnotesize $2.58 \pm 0.05$} & {\footnotesize $2.17 \pm 0.07$} & {\footnotesize $2.1 \pm 0.2$} & {\footnotesize $2.435 \pm 0.006$} & {\footnotesize $2.8 \pm 0.2$} & -- & -- \\
& & & & & & & & & & \\
{\footnotesize $\theta_{path}$} & {\footnotesize $0$} & -- & {\footnotesize $0.63 \pm 0.04$} & -- & {\footnotesize $0.28 \pm 0.02$} & -- & {\footnotesize $1.07 \pm 0.08$} & -- & {\footnotesize $1.23 \pm 0.07$} & -- \\
& & & & & & & & & & \\
{\footnotesize $t_{path}$} & {\footnotesize $2$} & {\footnotesize $2.04 \pm 0.04$} & {\footnotesize $4.2 \pm 0.1$} & {\footnotesize $4.6 \pm 0.1$} & {\footnotesize $2.7 \pm 0.1$} & {\footnotesize $2.46 \pm 0.04$} & {\footnotesize $3.8 \pm 0.1$} & {\footnotesize $3.9 \pm 0.3$} & {\footnotesize $3.8 \pm 0.2$} & {\footnotesize $7.7 \pm 1.8$} \\
& & & & & & & & & & \\
{\footnotesize $\theta_{tree}$} & {\footnotesize $-0.1 \pm 0.2$} & {\footnotesize $0$} & {\footnotesize $-0.14 \pm 0.02$} & {\footnotesize $0.08 \pm 0.09$} & {\footnotesize $-0.3 \pm 0.1$} & {\footnotesize $0.1 \pm 0.2$} & {\footnotesize $-0.96 \pm 0.02$} & {\footnotesize $-0.9 \pm 0.2$} & {\footnotesize $-0.84 \pm 0.07$} & {\footnotesize $-0.8 \pm 0.3$} \\
& & & & & & & & & & \\
{\footnotesize $t_{tree}$} & {\footnotesize $1.3 \pm 0.3$} & {\footnotesize $1.33 \pm 0.04$} & {\footnotesize $1.857 \pm 0.005$} & {\footnotesize $1.92 \pm 0.07$} & {\footnotesize $1.52 \pm 0.04$} & {\footnotesize $1.47 \pm 0.04$} & {\footnotesize $2.19 \pm 0.02$} & {\footnotesize $1.9 \pm 0.2$} & {\footnotesize $2.10 \pm 0.01$} & {\footnotesize $1.9 \pm 0.2$} \\
& & & & & & & & & & \\
{\footnotesize $\gamma_c$} & {\footnotesize $1.27 \pm 0.02$} & {\footnotesize $1.3 \pm 0.1$} & {\footnotesize $1.00 \pm 0.08$} & {\footnotesize $1.11 \pm 0.05$} & {\footnotesize $1.1 \pm 0.1$} & {\footnotesize $1.14 \pm 0.09$} & {\footnotesize $0.98 \pm 0.02$} & {\footnotesize $1.02 \pm 0.09$} & {\footnotesize $1.00 \pm 0.02$} & {\footnotesize $1.0 \pm 0.2$} \\
& & & & & & & & & & \\
{\footnotesize $\beta$} & -- & -- & {\footnotesize $0.628 \pm 0.002$} & {\footnotesize $0.74 \pm 0.07$} & {\footnotesize $0.881 \pm 0.003$} & {\footnotesize $0.95 \pm 0.07$} & -- & -- & -- & --\\
& & & & & & & & & & \\
\hline
\end{tabular}
\caption{
\label{tab:ExpSummary}
Critical exponents for distribution functions of lattice trees.
(Top panel)
Theoretical predictions by using the results of the Flory theory for $\nu_{path}$, $\rho$ and $\nu$~\cite{Rosa2016a,Rosa2016b}.
(Bottom)
Corresponding numerical results obtained by:
(a)
Extrapolation to $N \rightarrow \infty$ of ``effective'' exponents for trees of size $N$, see also Supplemental Tables~SI-IV 
for details.
(b)
Substitution of numerical asymptotic exponents (from Refs.~\cite{Rosa2016a,Rosa2016b} and this work) into the scaling relations summarized in the top.
Mean values and corresponding error bars have been rounded to the first significant decimal digit.
}
\end{table*}

\section{Summary and Conclusion}\label{sec:concls}

In the present article, we have pursued our investigation of the conformational statistics of various types of lattice trees with volume interactions~\cite{Rosa2016a,Everaers2016a,Rosa2016b}:
$2d$ and $3d$ melts of lattice trees with annealed connectivity as well as $3d$ self-avoiding lattice trees with annealed and with quenched ideal connectivity.
The well understood case of ideal, non-interacting lattice trees with annealed connectivity~\cite{ZimmStockmayer49,DeGennes1968} always serves as a useful reference. 
Here we have complemented the earlier analyses~\cite{Rosa2016a,Rosa2016b} of the average behaviour by reporting results for distribution functions for observables characterising tree conformations and connectivities.
In particular, we found that branch weight distributions follow a generalized Kramers relation, Eq.~(\ref{eq:Kramers_SplitProb}), with ${\mathcal Z}_n \sim c^n n^{\epsilon-2}$
and that path length and distance distributions are non-Gaussian and closely follow Redner-des Cloizeaux~\cite{Redner1980,DesCloizeauxBook} distributions
Eqs.~(\ref{eq:pl},\ref{eq:q_RdC_l}), (\ref{eq:pr_of_l},\ref{eq:q_RdC_path}) and (\ref{eq:pr},\ref{eq:q_RdC_tree})
of the type
\begin{eqnarray*}
p_N(\vec r) &=& 
   \left(\frac1{\langle r^2\rangle_N} \right)^{d/2} \ 
  q\left(\frac{\left|\vec r \right|}{ \sqrt{\langle r^2\rangle_N}}\right)\\
q(\vec x) &=&  C(\theta,t) \, |x|^\theta\  \exp \left( -(K(\theta,t)\, |x|)^t \right) 
\end{eqnarray*}
which are fully characterised by pairs of additional exponents, $\theta$ and $t$, summarised in Table~\ref{tab:ExpSummary}.
The various exponents $t$ describe the compressed exponential large distance (large path length) behavior of the distribution.
Our results suggest, that they obey generalized Fisher-Pincus~\cite{FisherSAWShape1966,PincusBlob1976} relations, Eqs.~(\ref{eq:tl}), (\ref{eq:tPath}), and (\ref{eq:tTree}).
The exponents $\theta$ characterize small distance (small path length) power law behavior and are hence related to contact probabilities (Eqs.~(\ref{eq:gamma_c theta_tree}) and~(\ref{eq:betaDef2})).
We have related them to each other and the other tree exponents (Eqs.~(\ref{eq:theta_l}) and (\ref{eq:ThetaTree})).
The only exception is the exponent $\theta_{path}$ for the small distance behavior of the path end-to-end distance distribution.
The situation is similar to the well-known case of linear self-avoiding walks,
where the corresponding exponent $\theta$ and the closure probability are related to the entropy exponent $\gamma$, which can not be predicted by Flory theory.
Additional work is required to corroborate the proposed relation~\cite{GrosbergSoftMatter2014} between the self-avoiding walk exponents and those characterising tree melts.

In conclusion, interacting randomly branching polymers exhibit an extremely rich behaviour and swell by a combination of modified branching and path stretching~\cite{GutinGrosberg93}.
As previously shown~\cite{Everaers2016a}, the average behaviour~\cite{Rosa2016a,Rosa2016b} in the various regimes and crossovers can be well described by a generalised Flory theory.
Our present results demonstrate, that this is not the case for distribution functions.
Nonetheless, their non-Gaussian functional form can be characterised by a small set of exponents, which are related to each other and the standard trees exponents.
The good agreement of the predicted relations with the numerical data suggests that we now dispose of a coherent framework for describing the connectivity and conformational statistics of interacting trees in a wide range of situations.



\clearpage

\setcounter{section}{0}
\setcounter{figure}{0}
\setcounter{table}{0}
\setcounter{equation}{0}

\renewcommand{\figurename}{Fig. S}
\renewcommand{\tablename}{Table S}

\tableofcontents

\clearpage

\section{Supplemental Figures}

\begin{figure*}
\includegraphics[width=\textwidth]{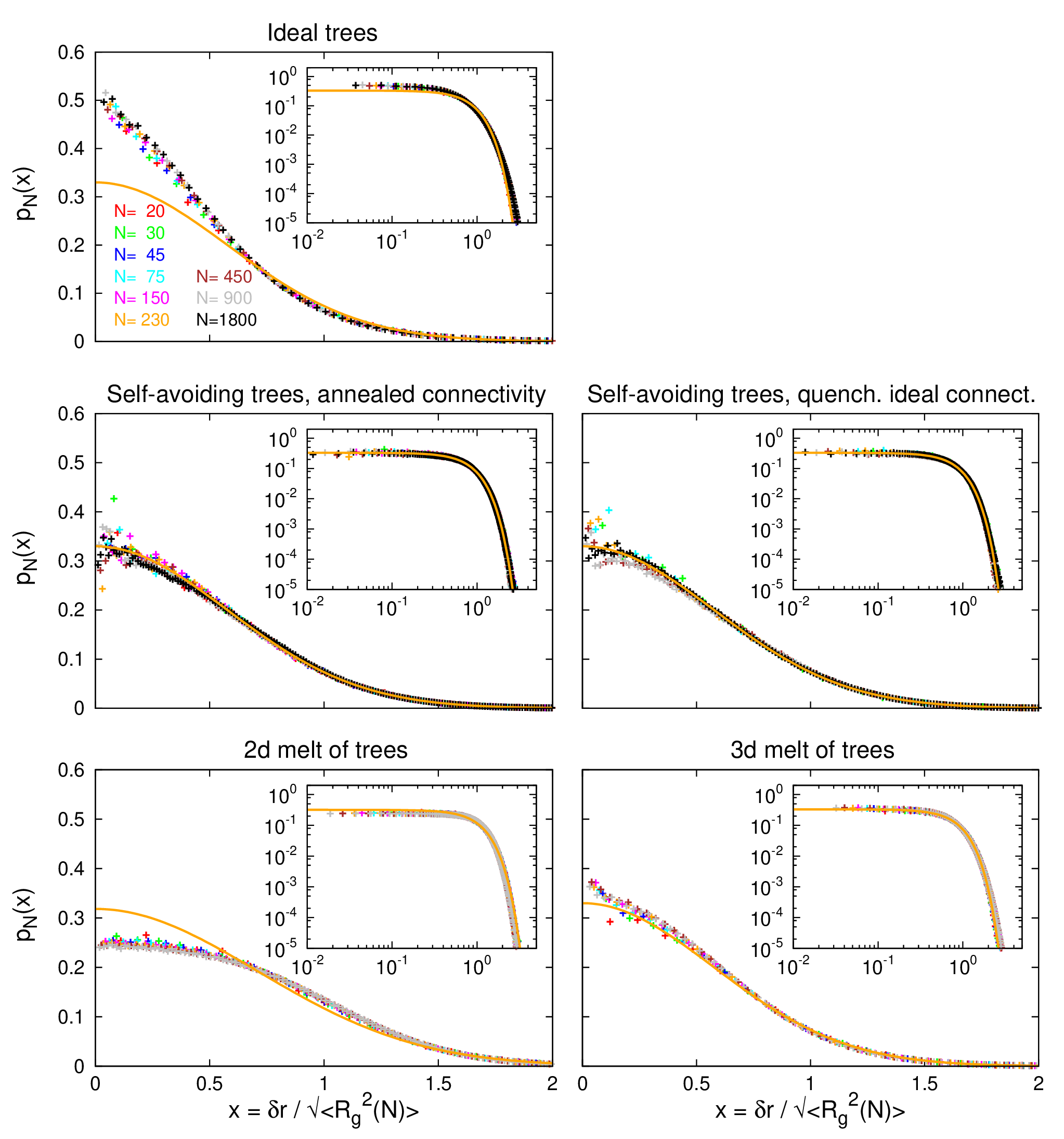}
\caption{
\label{fig:RCoM_Lcenter_Statistics}
Distribution functions of spatial distances, ${\vec{\delta r}}$, of nodes from the tree center of mass,
compared to the Gaussian distribution (yellow line).
For ideal trees, $p_N({\vec{\delta r}})$ is not Gaussian,
an effect similar to what reported a long time ago for ideal linear chains~\cite{DebyeBuecheJCP1952}.
Interestingly, for self-avoiding trees $p_N({\vec{\delta r}})$ becomes almost Gaussian.
Conversely, some noticeable deviations appear in melts of trees, especially in the $2d$ case.
In fact, given the node $i$ at spatial position from the center of mass ${\vec{\delta r}}_i$ and from the tree center $\vec{\delta_c r}_i$
one might be tempted to rationalise this behaviour by applying the convolution relation
$p_N({\vec{\delta r}}_i) = \int_0^{\infty} d {\vec{\delta r}_c} \, p_N({\vec{\delta r}_c}) \, p_N({\vec{\delta_c r}_i})$.
However, this would be incorrect as ${\vec{\delta r}_c}$ and ${\vec{\delta_c r}_i}$ are not statistically independent.
Insets show the same plots in log-log scale.
}
\end{figure*}

\clearpage

\section{Supplemental Tables}

\begin{table*}
\begin{tabular}{|c||c|c|c|c|}
\hline
{} & \multicolumn{4}{c|}{$\theta_{l}$} \\
\hline
\hline
& & & & \scs{Self-avoid. trees} \\
\scs{$N$} & \scs{Ideal trees} & \scs{$2d$ melt of trees} & \scs{$3d$ melt of trees} & \scs{anneal. connect.} \\
\hline
\scs{$20$} & \scs{$0.525 \pm 0.034$} & \scs{$0.390 \pm 0.026$} & \scs{$0.391 \pm 0.025$} & \scs{$0.365 \pm 0.038$} \\
\scs{$30$} & \scs{$0.537 \pm 0.019$} & \scs{$0.443 \pm 0.017$}  & \scs{$0.447 \pm 0.016$} & \scs{$0.365 \pm 0.019$} \\
\scs{$45$} & \scs{$0.563 \pm 0.012$} & \scs{$0.484 \pm 0.011$}  & \scs{$0.497 \pm 0.010$} & \scs{$0.391 \pm 0.011$} \\
\scs{$75$} & \scs{$0.608 \pm 0.009$} & \scs{$0.525 \pm 0.007$}  & \scs{$0.556 \pm 0.007$} & \scs{$0.409 \pm 0.006$} \\
\scs{$150$} & \scs{$0.677 \pm 0.007$} & \scs{$0.563 \pm 0.004$}  & \scs{$0.623 \pm 0.005$} & \scs{$0.443 \pm 0.003$} \\
\scs{$230$} & \scs{$0.722 \pm 0.007$} & \scs{$0.580 \pm 0.003$}  & \scs{$0.658 \pm 0.004$} & \scs{$0.466 \pm 0.003$} \\
\scs{$450$} & \scs{$0.778 \pm 0.005$} & \scs{$0.587 \pm 0.003$}  & \scs{$0.708 \pm 0.003$} & \scs{$0.471 \pm 0.002$} \\
\scs{$900$} & \scs{$0.831 \pm 0.004$} & \scs{$0.588 \pm 0.002$} & \scs{$0.751 \pm 0.002$} & \scs{$0.486 \pm 0.002$} \\
\scs{$1800$} & \scs{$0.875 \pm 0.003$} & & & \scs{$0.501 \pm 0.001$} \\
\hline
\scs{Best fit for $N\geq$} & \scs{$20$} & \scs{$20$} & \scs{$20$} & \scs{$20$} \\
\scs{$\Delta$} & \scs{$0.242 \pm 0.027$} & \scs{$1.110 \pm 0.081$} & \scs{$0.390 \pm 0.025$} & \scs{$0.374 \pm 0.001$} \\
\scs{${\tilde \chi}^2$} & \scs{$0.750$} & \scs{$1.164$} & \scs{$0.064$} & \scs{$3.786$} \\
\scs{$\mathcal Q$} & \scs{$0.610$} & \scs{$0.324$} & \scs{$0.997$} & \scs{$0.001$} \\
\scs{$N \rightarrow \infty$} & \scs{$1.188 \pm 0.075$} & \scs{$0.594 \pm 0.003$} & \scs{$0.905 \pm 0.031$} & \scs{$0.543 \pm 0.011$} \\
\hline
\scs{Best fit for $N\geq$} & \scs{$450$} & \scs{$230$} & \scs{$230$} & \scs{$450$} \\
\scs{$\Delta$} & \scs{1} & \scs{1} & \scs{1} & \scs{1} \\
\scs{${\tilde \chi}^2$} & \scs{$6.212$} & \scs{$0.281$} & \scs{$6.780$} & \scs{$7.274$} \\
\scs{$\mathcal Q$} & \scs{$0.013$} & \scs{$0.596$} & \scs{$0.009$} & \scs{$0.007$} \\
\scs{$N \rightarrow \infty$} & \scs{$0.909 \pm 0.004$} & \scs{$0.592 \pm 0.003$} & \scs{$0.785 \pm 0.004$} & \scs{$0.512 \pm 0.001$} \\
\hline
& \scs{$\mathbf{1.049}$} & \scs{$\mathbf{0.593}$} & \scs{$\mathbf{0.845}$} & \scs{$\mathbf{0.528}$} \\
& \scs{$\mathbf{\pm 0.075}$} & \scs{$\mathbf{\pm 0.003}$} & \scs{$\mathbf{\pm 0.031}$} & \scs{$\mathbf{\pm 0.011}$} \\
& \scs{$\mathbf{\pm 0.140}$} & \scs{$\mathbf{\pm 0.001}$} & \scs{$\mathbf{\pm 0.060}$} & \scs{$\mathbf{\pm 0.016}$} \\
\hline
\end{tabular}
\\
\begin{tabular}{|c||c|c|c|c|}
\multicolumn{5}{c}{}\\
\hline
{} & \multicolumn{4}{c|}{$t_{l}$} \\
\hline
\hline
& & & & \scs{Self-avoid. trees} \\
\scs{$N$} & \scs{Ideal trees} & \scs{$2d$ melt of trees} & \scs{$3d$ melt of trees} & \scs{anneal. connect.} \\
\hline
\scs{$20$} & \scs{$2.787 \pm 0.131$} & \scs{$3.325 \pm 0.152$} & \scs{$3.373 \pm 0.155$} & \scs{$2.394 \pm 0.135$} \\
\scs{$30$} & \scs{$2.742 \pm 0.074$} & \scs{$3.043 \pm 0.084$} & \scs{$3.098 \pm 0.083$} & \scs{$2.524 \pm 0.078$} \\
\scs{$45$} & \scs{$2.691 \pm 0.045$} & \scs{$2.842 \pm  0.049$} & \scs{$2.879 \pm 0.046$} & \scs{$2.519 \pm 0.045$} \\
\scs{$75$} & \scs{$2.599 \pm 0.030$} & \scs{$2.659 \pm 0.026$} & \scs{$2.699 \pm 0.028$} & \scs{$2.536 \pm 0.024$} \\
\scs{$150$} & \scs{$2.457 \pm 0.021$} & \scs{$2.513 \pm  0.014$} & \scs{$2.516 \pm 0.016$} & \scs{$2.464 \pm 0.012$} \\
\scs{$230$} & \scs{$2.384 \pm 0.018$} & \scs{$2.454 \pm 0.010$} & \scs{$2.441 \pm 0.013$} & \scs{$2.432 \pm 0.009$} \\
\scs{$450$} & \scs{$2.290 \pm 0.012$} & \scs{$2.405 \pm 0.008$} & \scs{$2.359 \pm 0.009$} & \scs{$2.422 \pm 0.006$} \\
\scs{$900$} & \scs{$2.212 \pm 0.008$} & \scs{$2.384 \pm 0.006$} & \scs{$2.278 \pm 0.006$} & \scs{$2.429 \pm 0.006$} \\
\scs{$1800$} & \scs{$2.169 \pm 0.005$} & & & \scs{$2.435 \pm 0.002$} \\
\hline
\scs{Best fit for $N\geq$} & \scs{$20$} & \scs{$20$} & \scs{$20$} & \scs{$20$} \\
\scs{$\Delta$} & \scs{$0.273 \pm 0.092$} & \scs{$0.853 \pm 0.141$} & \scs{$0.458 \pm 0.109$} & \scs{$1.110 \pm 1.208$} \\
\scs{${\tilde \chi}^2$} & \scs{$0.823$} & \scs{$0.264$} & \scs{$0.401$} & \scs{$3.316$} \\
\scs{$\mathcal Q$} & \scs{$0.552$} & \scs{$0.933$} & \scs{$0.848$} & \scs{$0.003$} \\
\scs{$N \rightarrow \infty$} & \scs{$1.911 \pm 0.075$} & \scs{$2.347 \pm 0.012$} & \scs{$2.105 \pm 0.040$} & \scs{$2.431 \pm 0.003$} \\
\hline
\scs{Best fit for $N\geq$} & \scs{$450$} & \scs{$230$} & \scs{$230$} & \scs{$450$} \\
\scs{$\Delta$} & \scs{1} & \scs{1} & \scs{1} & \scs{1} \\
\scs{${\tilde \chi}^2$} & \scs{$0.179$} & \scs{$0.059$} & \scs{$6.789$} & \scs{$0.025$} \\
\scs{$\mathcal Q$} & \scs{$0.672$} & \scs{$0.809$} & \scs{$0.009$} & \scs{$0.873$} \\
\scs{$N \rightarrow \infty$} & \scs{$2.130 \pm 0.008$} & \scs{$2.360 \pm 0.008$} & \scs{$2.228 \pm 0.008$} & \scs{$2.439 \pm 0.004$} \\
\hline
& \scs{$\mathbf{2.021}$} & \scs{$\mathbf{2.353}$} & \scs{$\mathbf{2.167}$} & \scs{$\mathbf{2.435}$} \\
& \scs{$\mathbf{\pm 0.075}$}& \scs{$\mathbf{\pm 0.012}$} & \scs{$\mathbf{\pm 0.040}$} & \scs{$\mathbf{\pm 0.004}$} \\
& \scs{$\mathbf{\pm 0.110}$} & \scs{$\mathbf{\pm 0.007}$} & \scs{$\mathbf{\pm 0.062}$} & \scs{$\mathbf{\pm 0.004}$} \\
\hline
\end{tabular}
\caption{
\label{tab:ThetalTl_Fits}
Path length statistics.
Effective exponents $\theta_l$ and $t_l$ obtained by best fits of the Redner-des Cloizeaux function, Eq.~(28) main paper. 
to the numerical distributions $p_N(l)$ of linear paths of length $l$ at different $N$ (Fig.~2, main paper). 
Extrapolations to $N \rightarrow \infty$ were obtained by employing
three- ($\Delta$ as free parameter)
and two-parameter ($\Delta$ fixed to $1$) fit functions,
see Sec.~IIIC main paper. 
Final estimates with corresponding statistical and systematic errors are given at the end of the table (in boldface).
}
\end{table*}

\begin{table*}
\begin{tabular}{|c|c||c|c||c|c||c|c||c|c|}
\hline
\multicolumn{2}{|c||}{} & \multicolumn{2}{c||}{} & \multicolumn{2}{c||}{} & \multicolumn{2}{c||}{$3d$ self-avoiding trees,} & \multicolumn{2}{c|}{$3d$ self-avoiding trees,} \\
\multicolumn{2}{|c||}{} & \multicolumn{2}{c||}{$2d$ melt of trees} & \multicolumn{2}{c||}{$3d$ melt of trees} & \multicolumn{2}{c||}{annealed connect.} & \multicolumn{2}{c|}{quenched {\it ideal} connect.} \\
\hline
\scs{$N$} & \scs{$l$} & \scs{$\theta_{path}$} & \scs{$t_{path}$} & \scs{$\theta_{path}$} & \scs{$t_{path}$} & \scs{$\theta_{path}$} & \scs{$t_{path}$} & \scs{$\theta_{path}$} & \scs{$t_{path}$} \\
\hline
\scs{$ 450$} & \scs{$16$} & \scs{$0.596 \pm 0.012$} & \scs{$4.098 \pm 0.045$} & \scs{$0.273 \pm 0.010$} & \scs{$2.749 \pm 0.017$} & \scs{$1.022 \pm 0.022$} & \scs{$3.833 \pm 0.045$} & \scs{$1.141 \pm 0.020$} & \scs{$3.672 \pm 0.047$} \\
\scs{$ 450$} & \scs{$32$} & \scs{$0.638 \pm 0.005$} & \scs{$4.179 \pm 0.021$} & \scs{$0.278 \pm 0.004$} & \scs{$2.521 \pm 0.006$} & \scs{$1.077 \pm 0.008$} & \scs{$3.698 \pm 0.017$} & \scs{$1.269 \pm 0.014$} & \scs{$3.835 \pm 0.032$} \\
\scs{$ 900$} & \scs{$16$} & \scs{$0.610 \pm 0.012$} & \scs{$4.127 \pm 0.048$} & \scs{$0.274 \pm 0.011$} & \scs{$2.775 \pm 0.018$} & \scs{$0.991 \pm 0.013$} & \scs{$3.934 \pm 0.028$} & \scs{$1.182 \pm 0.020$} & \scs{$3.587 \pm 0.044$} \\
\scs{$ 900$} & \scs{$32$} & \scs{$0.682 \pm 0.006$} & \scs{$4.324 \pm 0.023$} & \scs{$0.297 \pm 0.004$} & \scs{$2.552 \pm 0.006$} & \scs{$1.095 \pm 0.008$} & \scs{$3.743 \pm 0.015$} & \scs{$1.308 \pm 0.007$} & \scs{$3.901 \pm 0.016$} \\
\scs{$1800$} & \scs{$16$} & \scs{$ $} & \scs{$ $} & \scs{$ $} & \scs{$ $} & \scs{$0.994 \pm 0.015$} & \scs{$3.960 \pm 0.033$} & \scs{$1.165 \pm 0.019$} & \scs{$3.581 \pm 0.043$} \\
\scs{$1800$} & \scs{$32$} & \scs{$ $} & \scs{$ $} & \scs{$ $} & \scs{$ $} & \scs{$1.130 \pm 0.004$} & \scs{$3.786 \pm 0.008$} & \scs{$1.298 \pm 0.003$} & \scs{$3.902 \pm 0.007$} \\
\scs{$1800$} & \scs{$64$} & \scs{$ $} & \scs{$ $} & \scs{$ $} & \scs{$ $} & \scs{$1.210 \pm 0.004$} & \scs{$3.705 \pm 0.008$} & \scs{$1.259 \pm 0.008$} & \scs{$4.058 \pm 0.021$} \\
\hline
& & \scs{$\mathbf{0.631}$} & \scs{$\mathbf{4.182}$} & \scs{$\mathbf{0.281}$} & \scs{$\mathbf{2.649}$} & \scs{$\mathbf{1.074}$} & \scs{$\mathbf{3.809}$} & \scs{$\mathbf{1.232}$} & \scs{$\mathbf{3.791}$} \\
& & \scs{$\mathbf{\pm 0.012}$} & \scs{$\mathbf{\pm 0.048}$} & \scs{$\mathbf{\pm 0.011}$} & \scs{$\mathbf{\pm 0.018}$} & \scs{$\mathbf{\pm 0.022}$} & \scs{$\mathbf{\pm 0.045}$} & \scs{$\mathbf{\pm 0.020}$} & \scs{$\mathbf{\pm 0.047}$} \\
& & \scs{$\mathbf{\pm 0.033}$} & \scs{$\mathbf{\pm 0.087}$} & \scs{$\mathbf{\pm 0.010}$} & \scs{$\mathbf{\pm 0.114}$} & \scs{$\mathbf{\pm 0.074}$} & \scs{$\mathbf{\pm 0.098}$} & \scs{$\mathbf{\pm 0.063}$} & \scs{$\mathbf{\pm 0.168}$} \\
\hline
\end{tabular}
\caption{
\label{tab:ThetaPathTPath_Fits}
Conformational statistics of linear paths.
Effective exponents $\theta_{path}$ and $t_{path}$ obtained by best fits of the Redner-des Cloizeaux function, Eq.~(34) main paper. 
to the numerical distributions $p_N(r | l)$ of end-to-end spatial distances of linear paths of length $l$ at given $N$ (Fig.~3, main paper). 
Final estimates with statistical and systematic errors are given at the end of the table (in boldface).
}
\end{table*}

\begin{table*}
\begin{tabular}{|c||c|c|c|c|c|}
\hline
{} & \multicolumn{5}{c|}{{\footnotesize $\theta_{tree}$}} \\
\hline
\hline
& & & & {\footnotesize SA trees} & {\footnotesize SA trees} \\
{\footnotesize $N$} & {\footnotesize Ideal trees} & {\footnotesize $2d$ melt of trees} & {\footnotesize $3d$ melt of trees} & {\footnotesize ann. connect.} & {\footnotesize quen. {\it ideal} connect.} \\
\hline
{\scs $  20$} & {\scs $ 1.075 \pm 0.522$} & {\scs $0.424 \pm 0.142$} & {\scs $0.409 \pm 0.386$} & {\scs $-0.617 \pm 0.200$} & -- \\
{\scs $  30$} & {\scs $ 0.518 \pm 0.371$} & {\scs $0.027 \pm 0.059$} & {\scs $-0.333 \pm 0.156$} & {\scs $-0.964 \pm 0.088$} & {\scs $-0.824 \pm 0.020$} \\
{\scs $  45$} & {\scs $-0.279 \pm 0.166$} & {\scs $-0.072 \pm 0.035$} & {\scs $-0.547 \pm 0.082$} & {\scs $-1.082 \pm 0.042$} & -- \\
{\scs $  75$} & {\scs $-0.456 \pm 0.086$} & {\scs $-0.119 \pm 0.018$} & {\scs $-0.582 \pm 0.039$} & {\scs $-1.062 \pm 0.022$} & {\scs $-0.881 \pm 0.003$} \\
{\scs $ 150$} & {\scs $-0.453 \pm 0.041$} & {\scs $-0.167 \pm 0.004$} & {\scs $-0.500 \pm 0.017$} & {\scs $-1.031 \pm 0.010$} & -- \\
{\scs $ 230$} & {\scs $-0.420 \pm 0.026$} & {\scs $-0.157 \pm 0.003$} & {\scs $-0.449 \pm 0.011$} & {\scs $-1.012 \pm 0.007$} & {\scs $-0.869 \pm 0.004$} \\
{\scs $ 450$} & {\scs $-0.347 \pm 0.013$} & {\scs $-0.152 \pm 0.002$} & {\scs $-0.392 \pm 0.005$} & {\scs $-1.000 \pm 0.003$} & {\scs $-0.861 \pm 0.002$} \\
{\scs $ 900$} & {\scs $-0.272 \pm 0.007$} & {\scs $-0.144 \pm 0.001$} & {\scs $-0.347 \pm 0.002$} & {\scs $-0.986 \pm 0.002$} & {\scs $-0.857 \pm 0.002$} \\
{\scs $1800$} & {\scs $-0.225 \pm 0.004$} & {\scs $ $} & {\scs $ $} & {\scs $-0.978 \pm 0.001$} & {\scs $-0.842 \pm 0.002$} \\
\hline
\scs{Best fit for $N\geq$} & \scs{$150$} & \scs{$150$} & \scs{$150$} & \scs{$150$} & -- \\
{\scs $\Delta$} & {\scs $0.323 \pm 0.175$} & {\scs $0.609 \pm 0.379$} & {\scs $0.448 \pm 1.52$} & {\scs $0.492 \pm 2.112$} & -- \\
{\scs ${\tilde \chi}^2$} & {\scs $0.493$} & {\scs $1.444$} & {\scs $0.037$} & {\scs $0.230$} & -- \\
{\scs $\mathcal Q$} & {\scs $0.611$} & {\scs $0.230$} & {\scs $0.847$} & {\scs $0.794$} & -- \\
{\scs $N \rightarrow \infty$} & {\scs $-0.023  \pm 0.158$} & {\scs $-0.132 \pm 0.016$} & {\scs $-0.224 \pm 0.094$} & {\scs $-0.957 \pm 0.017$} & -- \\
\hline
\scs{Best fit for $N\geq$} & \scs{$450$} & \scs{$230$} & \scs{$230$} & \scs{$450$} & \scs{$450$} \\
{\scs $\Delta$} & {\scs 1} & {\scs 1} & {\scs 1} & {\scs 1} & {\scs 1} \\
{\scs ${\tilde \chi}^2$} & {\scs $0.454$} & {\scs $2.004$} & {\scs $2.425$} & {\scs $0.223$} & {\scs $16.245$} \\
{\scs $\mathcal Q$} & {\scs $0.501$} & {\scs $0.157$} & {\scs $0.119$} & {\scs $0.637$} & {\scs $6 \times 10^{-5}$} \\
{\scs $N \rightarrow \infty$} & {\scs $-0.184  \pm 0.007$} & {\scs $-0.140 \pm 0.002$} & {\scs $-0.310 \pm 0.005$} & {\scs $-0.971 \pm 0.002$} & {\scs $-0.842 \pm 0.003$} \\
\hline
& {\scs $\mathbf{-0.104}$} & {\scs $\mathbf{-0.136}$} & {\scs $\mathbf{-0.267}$} & {\scs $\mathbf{-0.964}$} & {\scs $\mathbf{-0.842}$} \\
& {\scs $\mathbf{\pm 0.158}$} & {\scs $\mathbf{\pm 0.016}$} & {\scs $\mathbf{\pm 0.094}$} & {\scs $\mathbf{\pm 0.017}$} & {\scs $\mathbf{\pm (0.060)}$} \\
& {\scs $\mathbf{\pm 0.081}$} & {\scs $\mathbf{\pm 0.004}$}  & {\scs $\mathbf{\pm 0.043}$} & {\scs $\mathbf{\pm 0.007}$} & {\scs $\mathbf{\pm (0.030)}$} \\
\hline
\multicolumn{6}{c}{} \\ 
\multicolumn{6}{c}{} \\
\end{tabular}
\begin{tabular}{|c||c|c|c|c|c|}
\hline
{} & \multicolumn{5}{c|}{{\footnotesize $t_{tree}$}} \\
\hline
\hline
& & & & {\footnotesize SA trees} & {\footnotesize SA trees} \\
{\footnotesize $N$} & {\footnotesize Ideal trees} & {\footnotesize $2d$ melt of trees} & {\footnotesize $3d$ melt of trees} & {\footnotesize ann. connect.} & {\footnotesize quen. {\it ideal} connect.} \\
\hline
{\scs $  20$} & {\scs $0.783 \pm 0.080$} & {\scs $1.253 \pm 0.065$} & {\scs $0.993 \pm 0.094$} & {\scs $1.244 \pm 0.095$} & -- \\
{\scs $  30$} & {\scs $0.900 \pm 0.078$} & {\scs $1.506 \pm 0.045$} & {\scs $1.306 \pm 0.073$} & {\scs $1.604 \pm 0.080$} & {\scs $1.816 \pm 0.022$} \\
{\scs $  45$} & {\scs $1.192 \pm 0.065$} & {\scs $1.617 \pm 0.034$} & {\scs $1.493 \pm 0.056$} & {\scs $1.845 \pm 0.060$} & -- \\
{\scs $  75$} & {\scs $1.335 \pm 0.047$} & {\scs $1.689 \pm 0.023$} & {\scs $1.608 \pm 0.034$} & {\scs $2.012 \pm 0.044$} & {\scs $2.005 \pm 0.005$} \\
{\scs $ 150$} & {\scs $1.401 \pm 0.027$} & {\scs $1.830 \pm 0.006$} & {\scs $1.597 \pm 0.017$} & {\scs $2.029 \pm 0.023$} & {\scs --} \\
{\scs $ 230$} & {\scs $1.410 \pm 0.019$} & {\scs $1.827 \pm 0.005$} & {\scs $1.574 \pm 0.011$} & {\scs $2.045 \pm 0.018$} & {\scs $2.104 \pm 0.010$} \\
{\scs $ 450$} & {\scs $1.393 \pm 0.010$} & {\scs $1.850 \pm 0.003$} & {\scs $1.553 \pm 0.005$} & {\scs $2.072 \pm 0.011$} & {\scs $2.096 \pm 0.007$} \\
{\scs $ 900$} & {\scs $1.354 \pm 0.005$} & {\scs $1.849 \pm 0.003$} & {\scs $1.543 \pm 0.003$} & {\scs $2.112 \pm 0.007$} & {\scs $2.080 \pm 0.006$} \\
{\scs $1800$} & {\scs $1.332 \pm 0.003$} & {\scs $ $} & {\scs $ $} & {\scs $2.159 \pm 0.004$} & {\scs $2.127 \pm 0.009$} \\
\hline
\scs{Best fit for $N\geq$} & \scs{$230$} & \scs{$20$} & \scs{$75$} & \scs{$20$} & \scs{$30$} \\
{\scs $\Delta$} & {\scs $0.240 \pm 4.265$} & {\scs $1.401 \pm 0.101$} & {\scs $0.563 \pm 0.432$} & {\scs $0.609 \pm 0.039$} & {\scs $1.224 \pm 0.110$} \\
{\scs ${\tilde \chi}^2$} & {\scs $0.961$} & {\scs $3.645$} & {\scs $0.302$} & {\scs $5.354$} & {\scs $7.785$} \\
{\scs $\mathcal Q$} & {\scs $0.327$} & {\scs $0.003$} & {\scs $ 0.739$} & {\scs $2 \times 10^{-5}$}& {\scs $3 \times 10^{-5}$} \\
{\scs $N \rightarrow \infty$} & {\scs $1.198 \pm 0.283$} & {\scs $1.855 \pm 0.003$} & {\scs $1.516 \pm 0.039$} & {\scs $2.197 \pm 0.016$} & {\scs $2.104 \pm 0.005$} \\
\hline
\scs{Best fit for $N\geq$} & \scs{$450$} & \scs{$230$} & \scs{$230$} & \scs{$450$} & \scs{$450$} \\
{\scs $\Delta$} & {\scs $1$} & {\scs $1$} & {\scs $1$} & {\scs $1$} & {\scs $1$} \\
{\scs ${\tilde \chi}^2$} & {\scs $0.058$} & {\scs $4.336$} & {\scs $0.004$} & {\scs $4.979$} & {\scs $16.411$} \\
{\scs $\mathcal Q$} & {\scs $0.809$} & {\scs $0.037$} & {\scs $0.948$} & {\scs $0.026$} & {\scs $5 \times 10^{-5}$} \\
{\scs $N \rightarrow \infty$} & {\scs $1.312 \pm 0.005$} & {\scs $1.859 \pm 0.004$} & {\scs $1.532 \pm 0.005$} & {\scs $2.187 \pm 0.007$} & {\scs $2.104 \pm 0.010$} \\
\hline
& {\scs $\mathbf{1.255}$} & {\scs $\mathbf{1.857}$} & {\scs $\mathbf{1.524}$} & {\scs $\mathbf{2.192}$} & {\scs $\mathbf{2.104}$} \\
& {\scs $\mathbf{\pm 0.283}$} & {\scs $\mathbf{\pm 0.004}$} & {\scs $\mathbf{\pm 0.039}$} & {\scs $\mathbf{\pm 0.016}$} & {\scs $\mathbf{\pm 0.010}$} \\
& {\scs $\mathbf{\pm 0.057}$} & {\scs $\mathbf{\pm 0.002}$} & {\scs $\mathbf{\pm 0.008}$} & {\scs $\mathbf{\pm 0.005}$} & {\scs $\mathbf{\pm 0.000}$} \\
\hline
\end{tabular}
\caption{
\label{tab:ThetaTreeTTree_Fits}
Conformational statistics of trees.
Effective exponents $\theta_{tree}$ and $t_{tree}$
obtained by best fits the Redner-des Cloizeaux distribution function, Eq.~(37) main paper. 
to the numerical distributions $p_N(r)$ of spatial distances $r$ between tree nodes at different $N$ (Fig.~4, main paper). 
Extrapolations to $N \rightarrow \infty$ were obtained by employing
three- ($\Delta$ as free parameter)
and two-parameter ($\Delta$ fixed to $1$) fit functions,
see Sec.~IIIC main paper. 
The three-parameter fit fails for $\theta_{tree}$ of self-avoiding trees with quenched ideal connectivity.
In this case, the reported statistical and systematic errors were based on the ones of ideal trees where:
(statistical error for three-parameter fit) $\approx$ 20(statistical error for two-parameter fit) $\approx$ 2(systematic error).
Final estimates with corresponding statistical and systematic errors are given at the end of the table (in boldface).
}
\end{table*}

\begin{table*}
\begin{tabular}{|c|c|c|c|c|c|c|c|}
\hline
& & \multicolumn{2}{c|}{} & \multicolumn{2}{c|}{} & {\footnotesize Self-avoid. trees,} & {\footnotesize Self-avoid. trees,} \\
& {\footnotesize Ideal trees} & \multicolumn{2}{c|}{\footnotesize $2d$ melt of trees} & \multicolumn{2}{c|}{\footnotesize $3d$ melt of trees} & {\footnotesize annealed connect.} & {\footnotesize quen. {\it ideal} connect.} \\
\hline
\hline
{\footnotesize $N$} & {\footnotesize $\langle N_c \rangle / (N+1)$} & {\footnotesize $\langle N_c \rangle / (N+1)$} & {\footnotesize $\langle N_c^{inter} \rangle / (N+1)$} & {\footnotesize $\langle N_c \rangle / (N+1)$} & {\footnotesize $\langle N_c^{inter} \rangle / (N+1)$} & {\footnotesize $\langle N_c \rangle / (N+1)$} & {\footnotesize $\langle N_c \rangle / (N+1)$} \\
\hline
\hline
{\footnotesize 3} & {\footnotesize $0.2284 \pm 0.0025$} & {\footnotesize $0.2865 \pm 0.0030$} & {\footnotesize $1.6775 \pm 0.0038$} & {\footnotesize $0.1923 \pm 0.0018$} & {\footnotesize $1.7675 \pm 0.0026$} & {\footnotesize $0.0000 \pm 0.0000$} & {\footnotesize $0.0210 \pm 0.0032$} \\
{\footnotesize 5} & {\footnotesize $0.3241 \pm 0.0040$} & {\footnotesize $0.3610 \pm 0.0027$} & {\footnotesize $1.2774 \pm 0.0033$} & {\footnotesize $0.2400 \pm 0.0016$} & {\footnotesize $1.4079 \pm 0.0022$} & {\footnotesize $0.0000 \pm 0.0000$} & {\footnotesize $0.0083 \pm 0.0017$} \\
{\footnotesize 10} & {\footnotesize $0.5306 \pm 0.0067$} & {\footnotesize $0.4890 \pm 0.0022$} & {\footnotesize $0.9235 \pm 0.0023$} & {\footnotesize $0.2958 \pm 0.0015$} & {\footnotesize $1.1255 \pm 0.0018$} & {\footnotesize $0.0146 \pm 0.0056$} & {\footnotesize $0.0062 \pm 0.0010$} \\
{\footnotesize 20} & {\footnotesize $0.7626 \pm 0.0026$} & {\footnotesize $0.5990 \pm 0.0017$} & {\footnotesize $0.6976 \pm 0.0017$} & {\footnotesize $0.3475 \pm 0.0009$} & {\footnotesize $0.9572 \pm 0.0011$} & {\footnotesize $0.0085 \pm 0.0009$} & \\
{\footnotesize 30} & {\footnotesize $0.9201 \pm 0.0027$} & {\footnotesize $0.6604 \pm 0.0014$} & {\footnotesize $0.5967 \pm 0.0013$} & {\footnotesize $0.3785 \pm 0.0008$} & {\footnotesize $0.8861 \pm 0.0008$} & {\footnotesize $0.0082 \pm 0.0007$} & {\footnotesize $0.0082 \pm 0.0007$} \\
{\footnotesize 45} & {\footnotesize $1.0893 \pm 0.0030$} & {\footnotesize $0.7190 \pm 0.0007$} & {\footnotesize $0.5122 \pm 0.0007$} & {\footnotesize $0.4088 \pm 0.0008$} & {\footnotesize $0.8291 \pm 0.0008$} & {\footnotesize $0.0064 \pm 0.0005$} & \\
{\footnotesize 75} & {\footnotesize $1.3286 \pm 0.0030$} & {\footnotesize $0.7861 \pm 0.0005$} & {\footnotesize $0.4238 \pm 0.0005$} & {\footnotesize $0.4471 \pm 0.0006$} & {\footnotesize $0.7703 \pm 0.0006$} & {\footnotesize $0.0073 \pm 0.0005$} & {\footnotesize $0.0091 \pm 0.0005$} \\
{\footnotesize 150} & {\footnotesize $1.6951 \pm 0.0032$} & {\footnotesize $0.8660 \pm 0.0004$} & {\footnotesize $0.3277 \pm 0.0004$} & {\footnotesize $0.4971 \pm 0.0005$} & {\footnotesize $0.7042 \pm 0.0005$} & {\footnotesize $0.0066 \pm 0.0003$} & \\
{\footnotesize 230} & {\footnotesize $1.9390 \pm 0.0034$} & {\footnotesize $0.9078 \pm 0.0004$} & {\footnotesize $0.2805 \pm 0.0003$} & {\footnotesize $0.5272 \pm 0.0004$} & {\footnotesize $0.6685 \pm 0.0004$} & {\footnotesize $0.0068 \pm 0.0002$} & {\footnotesize $0.0086 \pm 0.0003$} \\
{\footnotesize 450} & {\footnotesize $2.3936 \pm 0.0038$} & {\footnotesize $0.9648 \pm 0.0003$} & {\footnotesize $0.2181 \pm 0.0002$} & {\footnotesize $0.5726 \pm 0.0004$} & {\footnotesize $0.6177 \pm 0.0004$} & {\footnotesize $0.0070 \pm 0.0002$} & {\footnotesize $0.0084 \pm 0.0002$} \\
{\footnotesize 900} & {\footnotesize $2.9351 \pm 0.0061$} & {\footnotesize $1.0115 \pm 0.0004$} & {\footnotesize $0.1690 \pm 0.0003$} & {\footnotesize $0.6197 \pm 0.0006$} & {\footnotesize $0.5680 \pm 0.0006$} & {\footnotesize $0.0071 \pm 0.0001$} & {\footnotesize $0.0084 \pm 0.0001$} \\
{\footnotesize 1800} & {\footnotesize $3.5665 \pm 0.0100$} & & & & & {\footnotesize $0.0069 \pm 0.0001$} & {\footnotesize $0.0085 \pm 0.0001$} \\
\hline
\multicolumn{6}{c}{} \\
\hline
& {\footnotesize $\langle N_c \rangle \sim N^{\gamma_c}$} & {\footnotesize $\langle N_c \rangle \sim N^{\gamma_c}$} & {\footnotesize $\langle N_c^{inter} \rangle \sim N^{\beta}$} & {\footnotesize $\langle N_c \rangle \sim N^{\gamma_c}$} & {\footnotesize $\langle N_c^{inter} \rangle \sim N^{\beta}$} & {\footnotesize $\langle N_c \rangle \sim N^{\gamma_c}$} & {\footnotesize $\langle N_c \rangle \sim N^{\gamma_c}$} \\
\hline
{\footnotesize $\Delta$} & {\footnotesize $0.446 \pm 0.094$} & {\footnotesize $0.214 \pm 0.057$} & {\footnotesize $1.238 \pm 0.217$} & {\footnotesize $0.156 \pm 0.192$} & {\footnotesize $1.129 \pm 0.216$} & -- & -- \\
{\footnotesize ${\tilde \chi}^2$} & {\footnotesize $0.803$} & {\footnotesize $1.104$} & {\footnotesize $0.969$} & {\footnotesize $0.739$} & {\footnotesize $1.467$} & -- & -- \\
{\footnotesize ${\mathcal Q}$} & {\footnotesize $0.567$} & {\footnotesize $0.356$} & {\footnotesize $0.435$} & {\footnotesize $0.566$} & {\footnotesize $0.209$} & -- & -- \\
& {\footnotesize $\gamma_c = 1.253 \pm 0.010$} & {\footnotesize $\gamma_c = 0.931 \pm 0.018$} & {\footnotesize $\beta = 0.629 \pm 0.001$} & {\footnotesize $\gamma_c = 1.011 \pm 0.082$} & {\footnotesize $\beta = 0.884 \pm 0.001$} & -- & -- \\
\hline
{\footnotesize $\Delta$} & {\footnotesize $0$} & {\footnotesize $0$} & {\footnotesize $0$} & {\footnotesize $0$} & {\footnotesize $0$} & {\footnotesize $0$} & {\footnotesize $0$} \\
{\footnotesize ${\tilde \chi}^2$} & {\footnotesize $2.665$} & {\footnotesize $367.197$} & {\footnotesize $2.126$} & {\footnotesize $8.733$} & {\footnotesize $0.432$} & {\footnotesize $0.649$} & {\footnotesize $10^{-5}$} \\
{\footnotesize ${\mathcal Q}$} & {\footnotesize $0.103$} & {\footnotesize $<10^{-6}$} & {\footnotesize $0.119$} & {\footnotesize $0.003$} & {\footnotesize $0.511$} & {\footnotesize $0.421$} & {\footnotesize $0.998$} \\
& {\footnotesize $\gamma_c = 1.288 \pm 0.002$} & {\footnotesize $\gamma_c = 1.076 \pm 0.001$} & {\footnotesize $\beta = 0.626 \pm 0.001$} & {\footnotesize $\gamma_c = 1.116 \pm 0.001$} & {\footnotesize $\beta = 0.878 \pm 0.001$} & {\footnotesize $\gamma_c = 0.982 \pm 0.017$} & {\footnotesize $\gamma_c = 1.002 \pm 0.018$} \\
\hline
& {\footnotesize $\mathbf{\gamma_c = 1.271}$} & {\footnotesize $\mathbf{\gamma_c = 1.004}$} & {\footnotesize $\mathbf{\beta = 0.628}$} & {\footnotesize $\mathbf{\gamma_c = 1.064}$} & {\footnotesize $\mathbf{\beta = 0.881}$} & {\footnotesize $\mathbf{\gamma_c = 0.982}$} & {\footnotesize $\mathbf{\gamma_c = 1.002}$} \\
& {\footnotesize $\mathbf{\pm 0.010 \pm 0.018}$} & {\footnotesize $\mathbf{\pm 0.018 \pm 0.073}$} & {\footnotesize $\mathbf{\pm 0.001 \pm 0.002}$} & {\footnotesize $\mathbf{\pm 0.082 \pm 0.053}$} & {\footnotesize $\mathbf{\pm 0.001 \pm 0.003}$} & {\footnotesize $\mathbf{\pm 0.017}$} & {\footnotesize $\mathbf{\pm 0.018}$} \\
\hline
\end{tabular}
\caption{
\label{tab:Contacts}
(Top half of the table)
$\langle N_c \rangle / (N+1)$, average number of intra-chain contacts per node.
$\langle N_c^{inter} \rangle / (N+1)$, average number of inter-chain contacts per node in tree melts.
(Bottom half of the table)
Asymptotic ($N \rightarrow \infty$) estimation of corresponding critical exponents, $\gamma_c$ and $\beta$.
{\bf For ideal trees and melt of trees},
the numerical extrapolation scheme follows Ref.~\cite{MadrasJPhysA1992} and was the same adopted in our former works~\cite{Rosa2016a,Rosa2016b}.
It combines best fits to the data of:
(1) single power-law behavior ($\Delta=0$: $\log \langle N_c \rangle = c_1 + \gamma_c^{\Delta=0} \log N$, for data corresponding to the 3 largest $N$ of each set) and
(2) power-law behavior with a correction-to-scaling term ($\Delta \neq 0$: $\log \langle N_c \rangle = c_2 + c_3 N^{-\Delta} + \gamma_c^{\Delta\neq0} \log N$, for data with $N\gtrsim10$).
The reported values (last line of the table) are calculated as:
$\gamma_c = \frac{\gamma_c^{\Delta=0}+\gamma_c^{\Delta\neq0}}{2} \pm$ (largest statistical error) $\pm$ (spread between $\gamma_c^{\Delta=0}$ and $\gamma_c^{\Delta\neq0}$),
the last being an estimate for systematic errors due to finite-size effects~\cite{MadrasJPhysA1992}.
Equivalent expressions hold for $\langle N_c^{inter} \rangle$.
{\bf For self-avoiding trees},
finite-size effects appear completely negligible:
then, $\gamma_c$ was obtained by fitting data for $N\geq 450$ to just the single power-law function.
}
\end{table*}


\begin{thebibliography}{44}
\expandafter\ifx\csname natexlab\endcsname\relax\def\natexlab#1{#1}\fi
\expandafter\ifx\csname bibnamefont\endcsname\relax
  \def\bibnamefont#1{#1}\fi
\expandafter\ifx\csname bibfnamefont\endcsname\relax
  \def\bibfnamefont#1{#1}\fi
\expandafter\ifx\csname citenamefont\endcsname\relax
  \def\citenamefont#1{#1}\fi
\expandafter\ifx\csname url\endcsname\relax
  \def\url#1{\texttt{#1}}\fi
\expandafter\ifx\csname urlprefix\endcsname\relax\def\urlprefix{URL }\fi
\providecommand{\bibinfo}[2]{#2}
\providecommand{\eprint}[2][]{\url{#2}}

\bibitem[{\citenamefont{Isaacson and Lubensky}(1980)}]{IsaacsonLubensky}
\bibinfo{author}{\bibfnamefont{J.}~\bibnamefont{Isaacson}} \bibnamefont{and}
  \bibinfo{author}{\bibfnamefont{T.~C.} \bibnamefont{Lubensky}},
  \bibinfo{journal}{J. Physique Lett.} \textbf{\bibinfo{volume}{41}},
  \bibinfo{pages}{L469} (\bibinfo{year}{1980}).

\bibitem[{\citenamefont{Daoud and Joanny}(1981)}]{DaoudJoanny1981}
\bibinfo{author}{\bibfnamefont{M.}~\bibnamefont{Daoud}} \bibnamefont{and}
  \bibinfo{author}{\bibfnamefont{J.~F.} \bibnamefont{Joanny}},
  \bibinfo{journal}{J. Physique} \textbf{\bibinfo{volume}{42}},
  \bibinfo{pages}{1359} (\bibinfo{year}{1981}).

\bibitem[{\citenamefont{Gutin et~al.}(1993)\citenamefont{Gutin, Grosberg, and
  Shakhnovich}}]{GutinGrosberg93}
\bibinfo{author}{\bibfnamefont{A.~M.} \bibnamefont{Gutin}},
  \bibinfo{author}{\bibfnamefont{A.~Y.} \bibnamefont{Grosberg}},
  \bibnamefont{and} \bibinfo{author}{\bibfnamefont{E.~I.}
  \bibnamefont{Shakhnovich}}, \bibinfo{journal}{Macromolecules}
  \textbf{\bibinfo{volume}{26}}, \bibinfo{pages}{1293} (\bibinfo{year}{1993}).

\bibitem[{\citenamefont{Grosberg}(2014)}]{GrosbergSoftMatter2014}
\bibinfo{author}{\bibfnamefont{A.~Y.} \bibnamefont{Grosberg}},
  \bibinfo{journal}{Soft Matter} \textbf{\bibinfo{volume}{10}},
  \bibinfo{pages}{560} (\bibinfo{year}{2014}).

\bibitem[{\citenamefont{Everaers et~al.}(2016)\citenamefont{Everaers, Grosberg,
  Rubinstein, and Rosa}}]{Everaers2016a}
\bibinfo{author}{\bibfnamefont{R.}~\bibnamefont{Everaers}},
  \bibinfo{author}{\bibfnamefont{A.~Y.} \bibnamefont{Grosberg}},
  \bibinfo{author}{\bibfnamefont{M.}~\bibnamefont{Rubinstein}},
  \bibnamefont{and} \bibinfo{author}{\bibfnamefont{A.}~\bibnamefont{Rosa}},
  \bibinfo{journal}{In preparation}  (\bibinfo{year}{2016}).

\bibitem[{\citenamefont{Rosa and Everaers}(2016{\natexlab{a}})}]{Rosa2016a}
\bibinfo{author}{\bibfnamefont{A.}~\bibnamefont{Rosa}} \bibnamefont{and}
  \bibinfo{author}{\bibfnamefont{R.}~\bibnamefont{Everaers}},
  \bibinfo{journal}{J. Phys. A: Math. Gen.} \textbf{\bibinfo{volume}{49}},
  \bibinfo{pages}{345001} (\bibinfo{year}{2016}{\natexlab{a}}).

\bibitem[{\citenamefont{Rosa and Everaers}(2016{\natexlab{b}})}]{Rosa2016b}
\bibinfo{author}{\bibfnamefont{A.}~\bibnamefont{Rosa}} \bibnamefont{and}
  \bibinfo{author}{\bibfnamefont{R.}~\bibnamefont{Everaers}},
  \bibinfo{journal}{J. Chem. Phys., accepted. Preprint:
  http://arxiv.org/abs/1610.03100}  (\bibinfo{year}{2016}{\natexlab{b}}).

\bibitem[{\citenamefont{Janse~van Rensburg}(2015)}]{vanRensburgBook2015}
\bibinfo{author}{\bibfnamefont{E.~J.} \bibnamefont{Janse~van Rensburg}},
  \emph{\bibinfo{title}{The statistical mechanics of interacting walks,
  polygons, animals and vesicles}} (\bibinfo{publisher}{Oxford University
  Press}, \bibinfo{year}{2015}).

\bibitem[{\citenamefont{Bunde et~al.}(1995)\citenamefont{Bunde, Havlin, and
  Porto}}]{HavlinPRL1995}
\bibinfo{author}{\bibfnamefont{A.}~\bibnamefont{Bunde}},
  \bibinfo{author}{\bibfnamefont{S.}~\bibnamefont{Havlin}}, \bibnamefont{and}
  \bibinfo{author}{\bibfnamefont{M.}~\bibnamefont{Porto}},
  \bibinfo{journal}{Phys. Rev. Lett.} \textbf{\bibinfo{volume}{74}},
  \bibinfo{pages}{2714} (\bibinfo{year}{1995}).

\bibitem[{\citenamefont{Rubinstein and Colby}(2003)}]{RubinsteinColby}
\bibinfo{author}{\bibfnamefont{M.}~\bibnamefont{Rubinstein}} \bibnamefont{and}
  \bibinfo{author}{\bibfnamefont{R.~H.} \bibnamefont{Colby}},
  \emph{\bibinfo{title}{Polymer Physics}} (\bibinfo{publisher}{Oxford
  University Press}, \bibinfo{address}{New York}, \bibinfo{year}{2003}).

\bibitem[{\citenamefont{Khokhlov and Nechaev}(1985)}]{KhokhlovNechaev85}
\bibinfo{author}{\bibfnamefont{A.~R.} \bibnamefont{Khokhlov}} \bibnamefont{and}
  \bibinfo{author}{\bibfnamefont{S.~K.} \bibnamefont{Nechaev}},
  \bibinfo{journal}{Phys. Lett.} \textbf{\bibinfo{volume}{112A}},
  \bibinfo{pages}{156} (\bibinfo{year}{1985}).

\bibitem[{\citenamefont{Rubinstein}(1987)}]{RubinsteinRepton1987}
\bibinfo{author}{\bibfnamefont{M.}~\bibnamefont{Rubinstein}},
  \bibinfo{journal}{Phys. Rev. Lett.} \textbf{\bibinfo{volume}{59}},
  \bibinfo{pages}{1946} (\bibinfo{year}{1987}).

\bibitem[{\citenamefont{Obukhov et~al.}(1994)\citenamefont{Obukhov, Rubinstein,
  and Duke}}]{RubinsteinPRL1994}
\bibinfo{author}{\bibfnamefont{S.~P.} \bibnamefont{Obukhov}},
  \bibinfo{author}{\bibfnamefont{M.}~\bibnamefont{Rubinstein}},
  \bibnamefont{and} \bibinfo{author}{\bibfnamefont{T.}~\bibnamefont{Duke}},
  \bibinfo{journal}{Phys. Rev. Lett.} \textbf{\bibinfo{volume}{73}},
  \bibinfo{pages}{1263} (\bibinfo{year}{1994}).

\bibitem[{\citenamefont{Rosa and Everaers}(2014)}]{RosaEveraersPRL2014}
\bibinfo{author}{\bibfnamefont{A.}~\bibnamefont{Rosa}} \bibnamefont{and}
  \bibinfo{author}{\bibfnamefont{R.}~\bibnamefont{Everaers}},
  \bibinfo{journal}{Phys. Rev. Lett.} \textbf{\bibinfo{volume}{112}},
  \bibinfo{pages}{118302} (\bibinfo{year}{2014}).

\bibitem[{\citenamefont{Smrek and Grosberg}({2015})}]{SmrekGrosberg2015}
\bibinfo{author}{\bibfnamefont{J.}~\bibnamefont{Smrek}} \bibnamefont{and}
  \bibinfo{author}{\bibfnamefont{A.~Y.} \bibnamefont{Grosberg}},
  \bibinfo{journal}{{J. Phys.-Condes. Matter}} \textbf{\bibinfo{volume}{{27}}},
  \bibinfo{pages}{{064117}} (\bibinfo{year}{{2015}}).

\bibitem[{\citenamefont{Ge et~al.}(2016)\citenamefont{Ge, Panyukov, and
  Rubinstein}}]{RubinsteinMacromolecules2016}
\bibinfo{author}{\bibfnamefont{T.}~\bibnamefont{Ge}},
  \bibinfo{author}{\bibfnamefont{S.}~\bibnamefont{Panyukov}}, \bibnamefont{and}
  \bibinfo{author}{\bibfnamefont{M.}~\bibnamefont{Rubinstein}},
  \bibinfo{journal}{Macromolecules} \textbf{\bibinfo{volume}{49}},
  \bibinfo{pages}{708} (\bibinfo{year}{2016}).

\bibitem[{\citenamefont{Marko and Siggia}(1995)}]{MarkoSiggiaSuperCoiledDNA}
\bibinfo{author}{\bibfnamefont{J.}~\bibnamefont{Marko}} \bibnamefont{and}
  \bibinfo{author}{\bibfnamefont{E.}~\bibnamefont{Siggia}},
  \bibinfo{journal}{Phys. Rev. E} \textbf{\bibinfo{volume}{52}},
  \bibinfo{pages}{2912} (\bibinfo{year}{1995}).

\bibitem[{\citenamefont{Doi and Edwards}(1986)}]{DoiEdwards}
\bibinfo{author}{\bibfnamefont{M.}~\bibnamefont{Doi}} \bibnamefont{and}
  \bibinfo{author}{\bibfnamefont{S.~F.} \bibnamefont{Edwards}},
  \emph{\bibinfo{title}{The Theory of Polymer Dynamics}}
  (\bibinfo{publisher}{Oxford University Press}, \bibinfo{address}{New York},
  \bibinfo{year}{1986}).

\bibitem[{\citenamefont{{Janse van Rensburg} and
  Madras}(1992)}]{MadrasJPhysA1992}
\bibinfo{author}{\bibfnamefont{E.~J.} \bibnamefont{{Janse van Rensburg}}}
  \bibnamefont{and} \bibinfo{author}{\bibfnamefont{N.}~\bibnamefont{Madras}},
  \bibinfo{journal}{J. Phys. A: Math. Gen.} \textbf{\bibinfo{volume}{25}},
  \bibinfo{pages}{303} (\bibinfo{year}{1992}).

\bibitem[{\citenamefont{Zimm and Stockmayer}(1949)}]{ZimmStockmayer49}
\bibinfo{author}{\bibfnamefont{B.~H.} \bibnamefont{Zimm}} \bibnamefont{and}
  \bibinfo{author}{\bibfnamefont{W.~H.} \bibnamefont{Stockmayer}},
  \bibinfo{journal}{J. Chem. Phys.} \textbf{\bibinfo{volume}{17}},
  \bibinfo{pages}{1301} (\bibinfo{year}{1949}).

\bibitem[{\citenamefont{{De Gennes}}(1968)}]{DeGennes1968}
\bibinfo{author}{\bibfnamefont{P.-G.} \bibnamefont{{De Gennes}}},
  \bibinfo{journal}{Biopolymers} \textbf{\bibinfo{volume}{6}},
  \bibinfo{pages}{715} (\bibinfo{year}{1968}).

\bibitem[{\citenamefont{Parisi and Sourlas}(1981)}]{ParisiSourlasPRL1981}
\bibinfo{author}{\bibfnamefont{G.}~\bibnamefont{Parisi}} \bibnamefont{and}
  \bibinfo{author}{\bibfnamefont{N.}~\bibnamefont{Sourlas}},
  \bibinfo{journal}{Phys. Rev. Lett.} \textbf{\bibinfo{volume}{46}},
  \bibinfo{pages}{871} (\bibinfo{year}{1981}).

\bibitem[{\citenamefont{Hsu et~al.}(2005)\citenamefont{Hsu, Nadler, and
  Grassberger}}]{GrassbergerJPhysA2005}
\bibinfo{author}{\bibfnamefont{H.-P.} \bibnamefont{Hsu}},
  \bibinfo{author}{\bibfnamefont{W.}~\bibnamefont{Nadler}}, \bibnamefont{and}
  \bibinfo{author}{\bibfnamefont{P.}~\bibnamefont{Grassberger}},
  \bibinfo{journal}{J. Phys. A: Math. Gen.} \textbf{\bibinfo{volume}{38}},
  \bibinfo{pages}{775} (\bibinfo{year}{2005}).

\bibitem[{\citenamefont{Derrida and de~Seze}(1982)}]{DerridaDeseze1982}
\bibinfo{author}{\bibfnamefont{B.}~\bibnamefont{Derrida}} \bibnamefont{and}
  \bibinfo{author}{\bibfnamefont{L.}~\bibnamefont{de~Seze}},
  \bibinfo{journal}{J. Physique} \textbf{\bibinfo{volume}{43}},
  \bibinfo{pages}{475} (\bibinfo{year}{1982}).

\bibitem[{\citenamefont{Janssen and Stenull}(2011)}]{JanssenStenullPRE2011}
\bibinfo{author}{\bibfnamefont{H.-K.} \bibnamefont{Janssen}} \bibnamefont{and}
  \bibinfo{author}{\bibfnamefont{O.}~\bibnamefont{Stenull}},
  \bibinfo{journal}{Phys. Rev. E} \textbf{\bibinfo{volume}{83}},
  \bibinfo{pages}{051126} (\bibinfo{year}{2011}).

\bibitem[{\citenamefont{{De Gennes}}(1979)}]{DeGennesBook}
\bibinfo{author}{\bibfnamefont{P.-G.} \bibnamefont{{De Gennes}}},
  \emph{\bibinfo{title}{Scaling Concepts in Polymer Physics}}
  (\bibinfo{publisher}{Cornell University Press}, \bibinfo{address}{Ithaca},
  \bibinfo{year}{1979}).

\bibitem[{\citenamefont{des Cloizeaux and Jannink}(1989)}]{DesCloizeauxBook}
\bibinfo{author}{\bibfnamefont{J.}~\bibnamefont{des Cloizeaux}}
  \bibnamefont{and} \bibinfo{author}{\bibfnamefont{G.}~\bibnamefont{Jannink}},
  \emph{\bibinfo{title}{Polymers in Solution}} (\bibinfo{publisher}{Oxford
  University Press}, \bibinfo{address}{Oxford}, \bibinfo{year}{1989}).

\bibitem[{\citenamefont{Flory}(1953)}]{FloryChemBook}
\bibinfo{author}{\bibfnamefont{P.~J.} \bibnamefont{Flory}},
  \emph{\bibinfo{title}{Principles of Polymer Chemistry}}
  (\bibinfo{publisher}{Cornell University Press}, \bibinfo{address}{Ithaca
  (NY)}, \bibinfo{year}{1953}).

\bibitem[{\citenamefont{Grosberg and Nechaev}(2015)}]{GrosbergNechaev2015}
\bibinfo{author}{\bibfnamefont{A.~Y.} \bibnamefont{Grosberg}} \bibnamefont{and}
  \bibinfo{author}{\bibfnamefont{S.~K.} \bibnamefont{Nechaev}},
  \bibinfo{journal}{J. Phys. A-Math. Theor.} \textbf{\bibinfo{volume}{48}},
  \bibinfo{pages}{345003} (\bibinfo{year}{2015}).

\bibitem[{\citenamefont{Guttmann}(1987)}]{Guttmann1987}
\bibinfo{author}{\bibfnamefont{A.~J.} \bibnamefont{Guttmann}},
  \bibinfo{journal}{J. Phys. A: Math. Gen.} \textbf{\bibinfo{volume}{20}},
  \bibinfo{pages}{1839} (\bibinfo{year}{1987}).

\bibitem[{\citenamefont{Everaers et~al.}(1995)\citenamefont{Everaers, Graham,
  and Zuckermann}}]{EveraersJPA1995}
\bibinfo{author}{\bibfnamefont{R.}~\bibnamefont{Everaers}},
  \bibinfo{author}{\bibfnamefont{I.~S.} \bibnamefont{Graham}},
  \bibnamefont{and} \bibinfo{author}{\bibfnamefont{M.~J.}
  \bibnamefont{Zuckermann}}, \bibinfo{journal}{J. Phys. A: Math. Gen.}
  \textbf{\bibinfo{volume}{28}}, \bibinfo{pages}{1271} (\bibinfo{year}{1995}).

\bibitem[{\citenamefont{Redner}(1980)}]{Redner1980}
\bibinfo{author}{\bibfnamefont{S.}~\bibnamefont{Redner}}, \bibinfo{journal}{J.
  Phys. A: Math. Gen.} \textbf{\bibinfo{volume}{13}}, \bibinfo{pages}{3525}
  (\bibinfo{year}{1980}).

\bibitem[{\citenamefont{Fisher}(1966)}]{FisherSAWShape1966}
\bibinfo{author}{\bibfnamefont{M.~E.} \bibnamefont{Fisher}},
  \bibinfo{journal}{J. Chem. Phys.} \textbf{\bibinfo{volume}{44}},
  \bibinfo{pages}{616} (\bibinfo{year}{1966}).

\bibitem[{\citenamefont{Pincus}(1976)}]{PincusBlob1976}
\bibinfo{author}{\bibfnamefont{P.}~\bibnamefont{Pincus}},
  \bibinfo{journal}{Macromolecules} \textbf{\bibinfo{volume}{9}},
  \bibinfo{pages}{386} (\bibinfo{year}{1976}).

\bibitem[{\citenamefont{Lubensky and Isaacson}(1979)}]{LubenskyIsaacson1979}
\bibinfo{author}{\bibfnamefont{T.}~\bibnamefont{Lubensky}} \bibnamefont{and}
  \bibinfo{author}{\bibfnamefont{J.}~\bibnamefont{Isaacson}},
  \bibinfo{journal}{Phys. Rev. A} \textbf{\bibinfo{volume}{20}},
  \bibinfo{pages}{2130} (\bibinfo{year}{1979}).

\bibitem[{\citenamefont{Seitz and Klein}(1981)}]{SeitzKlein1981}
\bibinfo{author}{\bibfnamefont{W.~A.} \bibnamefont{Seitz}} \bibnamefont{and}
  \bibinfo{author}{\bibfnamefont{D.~J.} \bibnamefont{Klein}},
  \bibinfo{journal}{J. Chem. Phys.} \textbf{\bibinfo{volume}{75}},
  \bibinfo{pages}{5190} (\bibinfo{year}{1981}).

\bibitem[{\citenamefont{Duarte and Ruskin}(1981)}]{DuarteRuskin1981}
\bibinfo{author}{\bibfnamefont{J.~A. M.~S.} \bibnamefont{Duarte}}
  \bibnamefont{and} \bibinfo{author}{\bibfnamefont{H.~J.}
  \bibnamefont{Ruskin}}, \bibinfo{journal}{J. Physique}
  \textbf{\bibinfo{volume}{42}}, \bibinfo{pages}{1585} (\bibinfo{year}{1981}).

\bibitem[{\citenamefont{Fisher}(1978)}]{FisherPRL1978}
\bibinfo{author}{\bibfnamefont{M.~E.} \bibnamefont{Fisher}},
  \bibinfo{journal}{Phys. Rev. Lett.} \textbf{\bibinfo{volume}{40}},
  \bibinfo{pages}{1610} (\bibinfo{year}{1978}).

\bibitem[{\citenamefont{Kurtze and Fisher}(1979)}]{KurtzeFisherPRB1979}
\bibinfo{author}{\bibfnamefont{D.~A.} \bibnamefont{Kurtze}} \bibnamefont{and}
  \bibinfo{author}{\bibfnamefont{M.~E.} \bibnamefont{Fisher}},
  \bibinfo{journal}{Phys. Rev. B} \textbf{\bibinfo{volume}{20}},
  \bibinfo{pages}{2785} (\bibinfo{year}{1979}).

\bibitem[{\citenamefont{Bovier et~al.}(1984)\citenamefont{Bovier, Fr{\"o}hlich,
  and Glaus}}]{BovierFroelichGlaus1984}
\bibinfo{author}{\bibfnamefont{A.}~\bibnamefont{Bovier}},
  \bibinfo{author}{\bibfnamefont{J.}~\bibnamefont{Fr{\"o}hlich}},
  \bibnamefont{and} \bibinfo{author}{\bibfnamefont{U.}~\bibnamefont{Glaus}},
  \emph{\bibinfo{title}{``Branched Polymers and Dimensional Reduction'' in
  ``Critical Phenomena, Random Systems, Gauge Theories''}}
  (\bibinfo{publisher}{North-Holland, Amsterdam, K. Osterwalder and R. Stora
  (eds.)}, \bibinfo{year}{1984}).

\bibitem[{\citenamefont{Heermann et~al.}(1984)\citenamefont{Heermann, Hong, and
  Stanley}}]{StanleyJPhysA1984}
\bibinfo{author}{\bibfnamefont{H.~J.} \bibnamefont{Heermann}},
  \bibinfo{author}{\bibfnamefont{D.~C.} \bibnamefont{Hong}}, \bibnamefont{and}
  \bibinfo{author}{\bibfnamefont{H.~E.} \bibnamefont{Stanley}},
  \bibinfo{journal}{J. Phys. A: Math. Gen.} \textbf{\bibinfo{volume}{17}},
  \bibinfo{pages}{L261} (\bibinfo{year}{1984}).

\bibitem[{\citenamefont{Stauffer and Aharony}(1994)}]{StaufferAharonyBook}
\bibinfo{author}{\bibfnamefont{D.}~\bibnamefont{Stauffer}} \bibnamefont{and}
  \bibinfo{author}{\bibfnamefont{A.}~\bibnamefont{Aharony}},
  \emph{\bibinfo{title}{Introduction to percolation theory}}
  (\bibinfo{publisher}{Taylor \& Francis Inc.}, \bibinfo{year}{1994}).

\bibitem[{\citenamefont{Press et~al.}(1992)\citenamefont{Press, Teukolsky,
  Vetterling, and Flannery}}]{NumericalRecipes}
\bibinfo{author}{\bibfnamefont{W.~H.} \bibnamefont{Press}},
  \bibinfo{author}{\bibfnamefont{S.~A.} \bibnamefont{Teukolsky}},
  \bibinfo{author}{\bibfnamefont{W.~T.} \bibnamefont{Vetterling}},
  \bibnamefont{and} \bibinfo{author}{\bibfnamefont{B.~F.}
  \bibnamefont{Flannery}}, \emph{\bibinfo{title}{Numerical Recipes in Fortran}}
  (\bibinfo{publisher}{Cambridge University Press},
  \bibinfo{address}{Cambridge}, \bibinfo{year}{1992}), \bibinfo{edition}{2nd}
  ed.

\bibitem[{\citenamefont{Debye and Bueche}(1952)}]{DebyeBuecheJCP1952}
\bibinfo{author}{\bibfnamefont{P.}~\bibnamefont{Debye}} \bibnamefont{and}
  \bibinfo{author}{\bibfnamefont{F.}~\bibnamefont{Bueche}},
  \bibinfo{journal}{J. Chem. Phys.} \textbf{\bibinfo{volume}{20}},
  \bibinfo{pages}{1337} (\bibinfo{year}{1952}).

\end{thebibliography}

\end{document}